\documentclass[12pt,preprint]{aastex}
\usepackage{amssymb}
\usepackage{graphicx}
\usepackage{mathrsfs}

\begin{document}
\title{Hyperaccreting Disks around Magnetars for Gamma-Ray Bursts: Effects of Strong Magnetic Fields}
\author{Dong Zhang$^{1,2}$ and Z. G. Dai$^2$}
\affil{$^1$Department of Astronomy, Ohio State University, 140 W.
18th Ave., Columbus, OH 43210;
\\$^2$Department of Astronomy, Nanjing University, Nanjing 210093,
China \\\rm dzhang@astronomy.ohio-state.edu, dzg@nju.edu.cn}

\begin{abstract}
Hyperaccreting neutron star or magnetar disks cooled via neutrino
emission can be a candidate of gamma-ray burst (GRB) central
engines. The strong field $\geq10^{15}-10^{16}$ G of a magnetar can
play a significant role in affecting the disk properties and even
lead to the funnel accretion process. In this paper we investigate
the effects of strong fields on the disks around magnetars, and
discuss implications of such accreting magnetar systems for GRBs and
GRB-like events. We discuss quantum effects of the strong fields on
the disk thermodynamics and microphysics due to modifications of the
electron distribution and energy in the strong field environment,
and use the magnetohydrodynamical conservation equations to describe
the behavior of the disk flow coupled with a large scale field,
which is generated by the star-disk interaction. If the disk field
is open, the disk properties mainly depend on the ratio between
$|B_{\phi}/B_{z}|$ and $\Omega/\Omega_{K}$ with $B_{\phi}$ and
$B_{z}$ being the azimuthal and vertical components of the disk
field, $\Omega$ and $\Omega_{K}$ being the accretion flow angular
velocity and Keplerian velocity respectively. On the other hand, the
disk properties also depend on the magnetar spin period if the disk
field is closed. In general, stronger fields give higher disk
densities, pressures, temperatures and neutrino luminosity.
Moreover, strong fields will change the electron fraction and
degeneracy state significantly. A magnetized disk is always
viscously stable outside the Alfv\'{e}n radius, but will be
thermally unstable near the Alfv\'{e}n radius where the magnetic
field plays a more important role in transferring the angular
momentum and heating the disk than the viscous stress. The funnel
accretion process will be only important for an extremely strong
field, which creates a magnetosphere inside the Alfv\'{e}n radius
and truncates the plane disk. Because of higher temperature and more
concentrated neutrino emission of a ring-like belt region on the
magnetar surface covered by funnel accretion, the neutrino
annihilation rate from the accreting magnetar can be much higher
than that from an accreting neutron star without fields.
Furthermore, the neutrino annihilation mechanism which releases the
gravitational energy of the surrounding disk and the
magnetically-driven pulsar wind which extracts the stellar
rotational energy from the magnetar surface can work together to
generate and feed an ultra-relativistic jet along the stellar
magnetic poles.
\end{abstract}


\section{Introduction}

Hyperaccreting black hole systems formed in massive star collapses
or compact object binary mergers have been considered as a candidate
for gamma-ray burst (GRB) central engines for two decades (e.g.,
Eichler et al. 1989; Narayan et al. 1992; Woosley 1993). The
accreting systems can drive ultra or mildly relativistic jets with
huge energy via neutrino ($\nu\bar{\nu}$) annihilation process
(Popham et al. 1999) or magnetohydrodynamical (MHD) mechanism, such
as the Blandford-Znajek mechanism (Blandford \& Znajek 1977) or the
magnetorotational instability (MRI, Balbus \& Hawley 1991), and
finally produce GRB explosions. The neutrino annihilation efficiency
above hyperaccreting disks around black holes with the elaborate
considerations of disk geometry, rotation, and general relativity
effects have been studied by many authors (e.g., Ruffert et al.
1997, 1998; Popham et al. 1999; Asano \& Fukuyama 2000, 2001; Miller
et al. 2003; Birkl et al. 2007). Although the annihilation process
can provide sufficient total energy up to $10^{50}$ ergs s$^{-1}$
above neutrino-dominated flows with accretion rate $\sim 1M_{\odot}$
s$^{-1}$ (Gu et al. 2006; Liu et al. 2007), it is still under debate
whether such a process can successfully generate a relativistic jet
from the polar region of the central black hole. For example,
MacFadyen et al. (2001) discussed that the central black hole formed
in the ``collapsar scenario" is more frequent to be formed by
fallback process after a mild explosion (Type II collapsar) rather
than formed promptly (Type I collapsar). The Type II collapsar can
establish an accretion disk with accretion rate $\sim0.001-0.01
M_{\odot}$ s$^{-1}$, which is not sufficient to produce a jet by
neutrino annihilation. Numerical simulations showed that the MHD
mechanism may be more efficient to drive an energetic
magnetically-dominated wind and generate a GRB explosion
(Tchekhovskoy et al. 2008, 2009; Nagataki 2009), or the annihilation
process and MHD winds can work together to provide the energy for
GRB explosions (Harikae et al. 2009).

On the other hand, both observational and theoretical evidences show
that newborn neutron stars or magnetars rather than black holes can
form in the GRB central engines (e.g., Dai \& Lu 1998a, 1998b; Zhang
\& M\'{e}sz\'{a}ros 2001; Dai 2004; Dai et al. 2006; Mazzali et al.
2006; Soderberg et al. 2006; Shibata \& Taniguchi 2006; Lee \&
Ramirez-Ruiz 2007). Highly magnetized neutron stars formed by
accretion-induced-collapse, convection, $\alpha-\Omega$ dynamo
mechanism or differential rotation can extract their rotational
kinetic energy up to $\sim 10^{51}-10^{52}$ ergs by spinning down
via magnetic activities (e.g., Usov 1992; Duncan \& Thompson 1992;
Klu\'{z}niak \& Ruderman 1998). A thermally-driven neutrino wind is
dominated during the Kelvin-Helmholtz cooling epoch after the
neutron star formation, lasting from a few seconds to tens of
seconds based on the strengths of surface fields. After that,
magnetically-dominated or Poynting-dominated wind becomes
significant along the polar region (Wheeler et al. 2000; Thompson
2003; Thompson 2004; Metzger et al. 2007). The jets from magnetars
may be accelerated to higher Lorentz factor $\sim100-10^{3}$ at
large radius several tens of seconds after core bounce, and bring
its magnetic energy to kinetic energy, which is finally dissipated
in regular GRB internal shocks (Thompson et al. 2004; Metzger et al.
2007; Bucciantini et al. 2007, 2009; Lyubarsky 2009a, 2009b).

However, all the magnetized neutron star or magnetar models ignore
the accretion process which occurs onto a protoneutron star for the
first several seconds (Woosley \& Bloom 2006). In the scenario of
massive star collapse, the outgoing shock generated by a successful
Type Ibc supernova explosion with a velocity of $\sim10^{9}$ cm
s$^{-1}$ can evacuate a cavity around the new born compact star in
the center of a supernova remnant. However, whether the outgoing
shock in the core collapse of a massive star can dominate the
accretion process and turn the accretion around is still unknown. If
the rotational core collapse can lead to the formation of a neutron
star or a magnetar, there is possible that the prompt accretion or
fallback process can make a hyperaccreting disk around the young
formed star. On the other hand, recent simulations also showed the
possibility of a debris disk around a massive neutron star as the
outcome of a compact binary merger (e.g., Shibata \& Taniguchi 2006,
Lee \& Ramirez-Ruiz 2007). Similar to the black hole system, the
transiently existing torus around a neutron star or magnetar with
sufficient angular momentum can form a neutrino-cooled disk and
release its gravitational binding energy via neutrino emission.

The hyperaccreting neutron star or magnetar system is also proposed
as another possible central engine of GRBs. Zhang \& Dai (2008a,
2009, hereafter papers I and II) found that, due to the inner
surface boundary of the compact star, the disk has a denser, hotter
inner region with higher pressure compared to its black hole
counterpart. Also, the entire disk can cool more efficiently via
neutrino emission and produce a higher neutrino annihilation
luminosity. A heavily thermally-driven outflow from the surface of
the new-born neutron star at early times ($\sim 100$ms) prevent the
outflow from being accelerated to a high Lorentz factor (Dessart et
al. 2009). However, if the disk accretion rate and the neutrino
emission luminosity from the surface boundary layer are sufficiently
high, an energetic ultrarelativistic jet via neutrino annihilation
can be actually produced after the jet breaks out of the stellar
envelope around the disk-neutron star system.

Until now, none of the above works consider the effects of magnetic
fields on the hyperaccreting transient disks around the central
compact stars. The MHD mechanisms in the accreting black hole
systems mainly focus on energy extraction near the central black
holes, and the proto-magnetar magnetically-driven wind model
discusses wind generation and propagation along the polar regions.
In the hyperaccreting disk the spherical Alfv\'{e}n radius of the
central star can be estimated as
\begin{equation}
r_{A}\simeq0.207\dot{M}_{-2}^{-2/7}M_{1.4}^{-1/7}
\mu_{30}^{4/7} \textrm{km},
\end{equation}
where $\dot{M}=0.01\dot{M}_{-2}M_{\odot}$ s$^{-1}$ is the accretion
rate, $M=1.4M_{1.4}M_{\odot}$ is the central star mass,
$\mu=\mu_{30}10^{30}$ G cm$^{3}$ is the central magnetic flux.
Figure 1 shows the Alfv\'{e}n radius $r_{A}$ as a function of
accretion rate with various magnetic fields. If the surface field
$B_{0}$ is less than
\begin{equation}
B_{0}\leq B_{\rm crit}=0.89\times10^{15}\dot{M}_{-2}^{1/2}M_{1.4}^{1/4}
r_{*,6}^{-5/4} \textrm{G}
\end{equation}
with $r=r_{*,6}10^{6}$ cm being the star radius, the accretion flow
will continue to be confined in the disk plane without co-rotating
with the compact object or getting funneled onto the magnetar poles.
Papers I and II assumed the strength of stellar surface field is
less than $10^{15}$ G, and did not consider the effects of a
magnetic field in the disk. Recently, Xie et al. (2007, 2009) and
Lei et al. (2009) studied the structure of magnetized hyperaccreting
neutrino-cooled disks around black holes based on similar models,
while the magnetic fields are generated in the disks up to
$10^{15}-10^{16}$ G (see also Janiuk \& Yuan 2009). Xie et al.
(2007, 2009) showed that the magnetic braking and viscosity can
drive magnetically-dominated accretion flows rather than
neutrino-dominated, and turn the disk temperature to be lower than
that without field. Lei et al. (2009) investigated the properties of
the NDAF with the magnetic torque acted between the central black
hole and the disk. The neutrino annihilation luminosity can be
increased by one order of magnitude for accretion rate
$\sim0.5M_{\odot}$ s$^{-1}$, and the disk becomes thermally and
viscously unstable in the inner region. However, all of their works
neglected microphysics processes and the changes of equations of
state in strong magnetic fields, which may change the disk pressure
and various neutrino cooling rates in the neutrino-cooled disks
significantly. Furthermore, if we look into the neutron star disks,
we should keep in mind that the structure of magnetic fields in
disks around magnetized neutron stars or magnetars are very
different from their black hole counterparts. The origin of a
magnetic field in this case in the central compact star constructs
the initial topology of the strong field. An interaction between the
star and the surrounding disk makes the stellar magnetic field
partially thread the accretion disk.

Our motivation in this paper is to investigate the effects of a
strong magnetic field on the hyperaccreting disk around a magnetar
formed in the collapse of a massive star or the merger of a compact
object binary, and their observational phenomena related to GRBs and
associated events. The accretion process onto the magnetized neutron
star has been widely studied in X-ray binaries, T Tauri stars and
cataclysmic variables since Ghosh \& Lamb (1978, 1979a, 1979b), both
theoretically (e.g., K\"{o}nigl 1991; Ostriker and Shu 1995;
Lovelace et al. 1995; Wang 1995; Lai 1998), and numerically (e.g.,
Hayashi et al. 1996; Miller \& Stone 1997; Romanova et al. 2005,
2008; Long et al. 2007, 2008). The neutrino-cooled hyperaccreting
disks, on the other hand, are much hotter, denser with much higher
pressure and accretion rates compared to the normal disks. Therefore
the structure and radiation process in these magnetized disks must
be very different from the normal ones. For example, the size of the
magnetosphere near the central star can be ignored except for a
magnetic field $\geq10^{16}$ G. The pressure relation $B^{2}/8\pi\gg
P_{\rm matter}+\rho v_{r}^{2}$ will not be satisfied in most cases.
More attractively, we need to study the neutrino emission and
annihilation process which is almost the most important properties
of the hyperaccreting disks but never occur in the normal magnetized
ones.

This paper is organized as follows. In Section 2 we consider the
quantum effects of a strong magnetic field in the microphysical
scale on disk density, pressure and various neutrino cooling rates.
In Section 3, we discuss the disk conservation equations coupled
with the global fields both from the central magnetar and generated
via the magnetar-disk interaction. Also, we discuss the field
topology of the central magnetar. Combining the equations in Section
2 and 3, we discuss the properties of the hyperaccreting disks in
the strong magnetic fields numerically in Section 4. The structure
of the magnetized disks depends on the detailed field strength and
configuration. In Section 5.1, we further discuss the disk outflows,
which may provide the kinetic energy of the supernova associated
with a GRB (MacFadyen \& Woosley 1999; Kohri et al. 2005). Then in
Section 5.2, we describe the funnel process onto the magnetar
surface when the field strength is extremely high. The funnel
process will significantly affect the neutrino annihilation
efficiency. Section 6 is a broad discussion. We consider the
importance role played by Joule dissipation, nucleon reactions and
$r$-process occurring in the magnetized disk, as well as various
outcomes of the massive star collapse using a unified point of view.
Moreover, we particularly focus on the application of magnetized
hyperaccreting disks in the GRB and GRB-like events (e.g, X-ray
flashes), and use a unified point of view to discuss the core
collapse models related to GRBs. Conclusions are presented in
Section 7.

\section{Thermodynamics and Microphysics in Strong Magnetic Fields}
\subsection{Density and Pressure in Magnetic Fields}

The distribution and energy of charged particles will change
significantly if the magnetic field is greater than the critical
value $B_{ci}=m_{i}^{2}c^{3}/q_{i}\hbar$, where $m_{i}$ and $q_{i}$
are the mass and charge of the particle respectively. As the
critical value for electrons is $B_{ce}\simeq4.4\times10^{13}$ G and
$B_{cp}\simeq1.5\times10^{20}$ G, we have to adopt the relativistic
Dirac equation for electrons in the environment near a magnetar with
the field $B>4.4\times10^{13}$ G, while the protons can be still
considered as classical. The main modification of electron
distribution is that, the phase space factor in the absence of the
magnetic field should be replaced by the summation over possible
Landau level
\begin{equation}
\frac{2}{h^{3}}\int d^{3}p\longrightarrow
\sum_{n=0}^{\infty}g_{nL}\int \frac{eB_{m}}{h^{2}c}dp,\label{change}
\end{equation}
where $B_{m}$ is the field strength, and $g_{nL}=2-\delta_{n0}$ with
$\delta_{n0}$ being the Kronecker delta function.

The electron energies are given by (Johnson \& Lippmann 1949)
\begin{equation}
E_{e}=\sqrt{p^{2}c^{2}+m_{e}^{2}c^{4}+2n_{L}eB_{m}\hbar
c}=m_{e}c^{2}\sqrt{x^{2}+1+n_{L}b},\label{energy00}
\end{equation}
where $n$ labels the Landau level, $b=2e\hbar B_{m}/m_{e}^{2}c^{3}$
is the dimensionless magnetic field parameter, and $x=p/m_{e}c$ is
the dimensionless momentum.

The number density of electrons and positrons with the magnetic
field strength $B_{m}$ is
\begin{eqnarray}
n_{e^{\mp}}&=&\sum_{n=0}^{\infty}g_{nL}\int_{-\infty}^{\infty}dp\frac{eB_{m}}{h^{2}c}f_{e^{\mp}}\nonumber
\\&&=2\pi\left(\frac{m_{e}c}{h}\right)^{3}b\sum_{n=0}^{\infty}g_{nL}\int_{0}^{\infty}dxf_{e^{\mp}},\label{den01}
\end{eqnarray}
where
\begin{equation}
f_{e^{\mp}}=\frac{1}{e^{m_{e}c^{2}\sqrt{x^{2}+1+bn_{L}}/k_{B}T\mp\eta_{e}}+1}
\end{equation}
is the Fermi-Dirac function. The total pressure in the disk is
contributed by five terms, i.e., the pressure of nucleons, electrons
(including positrons), radiation, neutrino and magnetic field:
\begin{equation}
P=P_{\rm nuc}+P_{\rm rad}+P_{e}+P_{\nu}+P_{B}.
\end{equation}
We take the approximation that $P_{\rm nuc}$ and $P_{\rm rad}$ do
not change with magnetic fields\footnote{We neglect the
photodisintegration process which happens far from the central star
$r\geq400$ km, thus the nucleons are mainly composed of classical
protons and neutrons without $\alpha$-particles in the disk region
we focus on. The equations of state for protons and neutrons do not
change with field $B<10^{20}$ G as mentioned above.}, and
$P_{\nu}=u_{\nu}/3$ is only noticeable in very opaque regions in the
disk, where $u_{\nu}$ is the energy density of neutrinos. The
pressures of electrons and positrons in the strong magnetic field
are
\begin{eqnarray}
P_{e_{\mp}}&=&\sum_{n=0}^{\infty}g_{nL}\int_{-\infty}^{\infty}dp\frac{eB_{m}}{3h^{2}}
\frac{p^{2}c}{\sqrt{p^{2}c^{2}+m_{e}^{2}c^{4}+2n_{L}eB_{m}\hbar
c}}f_{e^{\mp}}\nonumber\\
&&=\frac{2\pi}{h^{3}}m_{e}^{4}c^{5}b\sum_{n=0}^{\infty}g_{nL}\int_{0}^{\infty}dx\frac{x^{2}}{\sqrt{x^{2}+1+n_{L}\cdot
b}}f_{e^{\mp}},\label{pre01}
\end{eqnarray}
and the electron pressure $P_{e}=P_{e^{-}}+P_{e^{+}}$. The pressure
of the magnetic field is $P_{B}=B_{m}^{2}/8\pi$.

If $B\rightarrow0$, the number density formula (\ref{den01}) and
(\ref{pre01}) as a series of different Landau levels will switch
back to their integral form without fields. However, the convergence
performance of Landau level series becomes poor for weak magnetic
fields. Figure 2 shows the convergence performance of density and
pressure Landau level series, and compares the summations of formula
(\ref{den01}) and (\ref{pre01}) to the values without fields. Here
we take temperature and chemical potential as fixed typical values
in neutrino-cooled disks. We define $b_{n}$ as the $n$ term in the
Landau level series of (\ref{den01}) and (\ref{pre01}). We note that
the value of $b_{n}$ drops faster for stronger magnetic fields. So
we have to take a huge number of $b_{n}$ terms to calculate the
summation for relatively weak magnetic fields. Moreover, whether the
microphysics change in the strong fields is important also depends
on the disk temperature. The difference between density or pressure
with and without fields is more significant for lower temperature.
This conclusion can also be applied to the hyperaccreting disks
around black holes which generate the fields by MRI or different
rotation. Later in this paper, we take the upper limit $n=10^{4}$
for $b_{n}$ summation calculation for $B\geq B_{ce}$, as the
pressure or the equations of state in a magnetized disk switch back
to those without fields for $B< B_{ce}$ and the typical temperature
in the disks.

\subsection{Neutrino Cooling in Magnetic Fields}
Neutrino cooling processes (especially, the Urca process) in the
environment of compact object magnetic fields have been studied for
years (e.g., Chen et al. 1974; Dorofeev et al. 1985; Dai et al.
1993; Yuan \& Zhang 1998; Lai \& Qian 1998; Roulet 1998; Baiko \&
Yakovlev 1999; Yakovlev et al. 2001; Duan \& Qian 2004, 2005; Luo
2005; Riquelme et al. 2005). Similarly, strong magnetic fields also
affect the neutrino emission in a hyperaccreting disk around the
compact object. Here we consider the problem systematically in this
section. The total neutrino cooling rates in the
vertically-integrated disk are taken as (Di Matteo et al. 2002;
Kohri et al. 2005)
\begin{equation}
Q_{\nu}^{-}=\sum_{i=e,\mu,\tau}\frac{(7/8)\sigma_{B}T^{4}}{(3/4)[\tau_{\nu_{i}}/2+1/\sqrt{3}+1/(3\tau_{a,\nu_{i}})]}
,\label{coolingrate1}
\end{equation}
where the total optical depth for three types of neutrinos
$\tau_{\nu_{i}}=\tau_{a,\nu_{i}}+\tau_{s,\nu_{i}}$ with
$\tau_{a,\nu_{i}}$ being the absorption depth and $\tau_{s,\nu_{i}}$
the scattering depth, both of which can be affected by strong
magnetic fields.

The neutrino absorption depth can be approximately given by (Popham
et a. 1999)
$\tau_{a,\nu_{i}}=\dot{q}_{a,\nu_{i}}H/(\frac{7}{2}\sigma_{B}T^{4})$,
where $\dot{q}_{a,\nu_{i}}$ are the absorption neutrino cooling
rates for three types of neutrino, and $H$ is the half thickness of
the disk. The electron neutrino cooling rate $\dot{q}_{a,\nu_{e}}$
in the disk can be simply taken as the summation of four terms
$\dot{q}_{a,\nu_{e}}=(\dot{q}_{eN}+\dot{q}_{e^{-}e^{+}\rightarrow\nu_{e}\bar{\nu}_{e}
}+\dot{q}_{\rm brems}+\dot{q}_{\rm plamson})H$, which are the the
cooling rates due to electron-positron capture by nucleons,
electron-positron pair annihilation into neutrinos, nucleon-nucleon
bremsstrahlung, and plasmon decays respectively (Kohri \& Mineshige
2002). On the other hand, the muon and tau neutrino cooling rates
are the summations of pair annihilation and bremsstrahlung
$\dot{q}_{a,\nu_{\mu}}H=\dot{q}_{a,\nu_{\tau}}H\simeq(\dot{q}_{e^{-}e^{+}\rightarrow\nu_{\tau}\bar{\nu}_{\tau}
}+\dot{q}_{\rm brems})H$.

The neutrino cooling via electron-positron capture by nucleons (or
say the Urca process in the disk), which is usually the most
significant cooling rate among various cooling terms in
hyperaccreting disks, can be considered as the combination of three
processes: $\dot{q}_{eN}=\dot{q}_{p+e^{-}\rightarrow
n+\nu_{e}}+\dot{q}_{n+e^{+}\rightarrow
p+\bar{\nu}_{e}}+\dot{q}_{n\rightarrow p+e^{-}+\bar{\nu}_{e}}$. If
we take the parameter $\tilde{K}$ as
\begin{equation}
\tilde{K}=\frac{G_{F}^{2}C_{V}^{2}(1+3g_{A}^{2})}{\pi^{3}\hbar^{7}}(m_{e}c)^{6},
\end{equation}
where $G_{F}^{2}\simeq1.436\times10^{-49}$ ergs cm$^{-3}$,
$C_{V}=1/2+2\textrm{sin}^{2}\theta_{W}$ with
$\textrm{sin}^{2}\theta_{W}=0.23$, and $g_{A}=-1.23$. Using the
modification rule in strong fields (\ref{change}), we can obtain the
three terms of neutrino cooling rates in strong magnetic fields as

\begin{equation}
\dot{q}_{p+e^{-}\rightarrow
n+\nu_{e}}=b\frac{\tilde{K}}{4}\sum_{n=0}^{\infty}g_{nL}n_{p}\int_{max\{q,\sqrt{1+n_{L}b}\}}^{\infty}
\frac{\varepsilon(\varepsilon-q)^{3}d\varepsilon}{\sqrt{\varepsilon^{2}-1-n_{L}b}}f_{e^{-}},\label{coolingrate3}
\end{equation}

\begin{equation}
\dot{q}_{n+e^{+}\rightarrow
p+\bar{\nu}_{e}}=b\frac{\tilde{K}}{4}\sum_{n=0}^{\infty}g_{nL}n_{n}\int_{\sqrt{1+n_{L}b}}^{\infty}
\frac{\varepsilon(\varepsilon+q)^{3}d\varepsilon}{\sqrt{\varepsilon^{2}-1-n_{L}b}}f_{e^{+}},\label{coolingrate4}
\end{equation}

\begin{equation}
\dot{q}_{n\rightarrow
p+e^{-}+\bar{\nu}_{e}}=b\frac{\tilde{K}}{4}\sum_{n=0}^{\infty}g_{nL}n_{n}\int_{\sqrt{1+n_{L}b}}^{q}
\frac{\varepsilon(q-\varepsilon)^{3}d\varepsilon}{\sqrt{\varepsilon^{2}-1-n_{L}b}}(1-f_{e^{-}}),\label{coolingrate5}
\end{equation}
where $\varepsilon=E_{\nu_{e}}/m_{e}c^{2}$ is the dimensionless
energy of electrons, and $q=(m_{n}-m_{p})/m_{e}\approx2.531$. Here
we adopt the approximation adopted in Shapiro \& Teukolsky (1983)
that the total reaction on nucleons is directly proportional to the
density number of nucleons, i.e., protons and neutrons. Figure 3
shows the convergence performance of Landau level series of two
dominated cooling rates (\ref{coolingrate3}) and
(\ref{coolingrate4}) with the typical temperature and chemical
potential in neutrino-cooled hyperaccreting disks. Similar to the
density and pressure in a strong magnetic field, the convergence
properties of neutrino cooling series also go poorer for a weaker
field. Moreover, the total Urca neutrino cooling rate becomes lower
for a higher magnetic field. Although the difference is not
significant for $T=5\times10^{10}$ K, it can be greater for a lower
temperature, as shown in Figure 2. We will discuss this effect later
in Sections 4 and 6.

The electron-positron pair annihilation rate without magnetic fields
can be calculated as (Burrows \& Thompson 2004)
\begin{equation}
\dot{q}_{e^{-}e^{+}\rightarrow\nu_{e}\bar{\nu}_{e}
}\simeq\dot{q}_{\nu_{e}\bar{\nu}_{e}(B=0)}\simeq
2.558\times10^{33}T_{11}^{9}f(\eta_{e})\rm\quad ergs\,cm^{-3}\,s^{-1},
\end{equation}
\begin{equation}
\dot{q}_{e^{-}e^{+}\rightarrow\nu_{\mu}\bar{\nu}_{\mu}
}=\dot{q}_{e^{-}e^{+}\rightarrow\nu_{\tau}\bar{\nu}_{\tau}
}\simeq
1.093\times10^{33}T_{11}^{9}f(\eta_{e})\rm\quad ergs\,cm^{-3}\,s^{-1},
\end{equation}
where the function $f(\eta_{e})$ is
\begin{equation}
f(\eta_{e})=\frac{F_{4}(\eta_{e})F_{3}(-\eta_{e})+F_{4}(-\eta_{e})F_{3}(\eta_{e})}{2F_{4}(0)F_{3}(0)}
\end{equation}
with the function $F_{n}(\pm\eta_{e})$ being
\begin{equation}
F_{n}(\pm\eta_{e})=\int_{0}^{\infty}\frac{x^{n}}{e^{x\mp\eta_{e}}+1}dx.
\end{equation}
In the case of strong magnetic fields, we can approximately take the
annihilation rate as
\begin{equation}
\dot{q}_{\nu_{i}\bar{\nu}_{i}}=\dot{q}_{\nu_{i}\bar{\nu}_{i}(B=0)}\times\frac{B_{m}e\hbar m_{e}^{2}c^{5}}{28\pi^{2}T^{4}k_{B}^{4}},
\end{equation}
when the field strength $B_{m}$ satisfies $B_{m}\hbar ec\gg
T^{2}k_{B}^{2}$ (e.g., Kaminker et al. 1992; Yakovlev et al. 2001).

In most cases the cooling processes via bremsstrahlung and plasmon
decay are much less significant than $e^{-}e^{+}$ capture and
annihilation, furthermore the unchanged distribution and energy of
neutrons and photons. Therefore we still use the formulae as in
Kohri et al. (2005) for these two processes.

Besides the absorption depth, three types of optical depth for
neutrinos through scattering off nucleons and electrons are given by
\begin{eqnarray}
\tau_{s,\nu_{i}} &= &(\sigma_{\nu_{i}p}n_{p}+\sigma_{\nu_{i}n}n_{n}+\sigma_{\nu_{i}e}n_{e})H\nonumber\\
 &&= (\sigma_{\nu_{i}p}Y_{p}+\sigma_{\nu_{i}n}Y_{n}+\sigma_{\nu_{i}e}Y_{e})\rho H/m_{B},
\end{eqnarray}
where $\sigma_{\nu_{i}p}$, $\sigma_{\nu_{i}n}$ and
$\sigma_{\nu_{i}e}$ are the cross sections of scattering on protons,
neutrons and electrons, $\rho$ is the density of the disk. We take
the cross sections of $\sigma_{\nu_{i}p}$ and $\sigma_{\nu_{i}n}$ in
strong magnetic fields as
\begin{equation}
\sigma_{\nu_{i}p}\simeq\sigma_{\nu_{i}p(B=0)}=\frac{\sigma_{0}}{6}\left(\frac{\varepsilon_{\nu}}{m_{e}c^{2}}\right)^{2}[(C_{V}-1)^{2}+5g_{A}^{2}(C_{A}-1)^{2}],
\end{equation}
\begin{equation}
\sigma_{\nu_{i}n}\simeq\sigma_{\nu_{i}n(B=0)}=\frac{\sigma_{0}}{4}\left(\frac{\varepsilon_{\nu}}{m_{e}c^{2}}\right)^{2}\left(\frac{1+5g_{A}^{2}}{6}\right),
\end{equation}
which are still classical (Tubbs \& Schramm 1975; Burrows \&
Thompson 2004). Here $\sigma_{0}\simeq1.705\times10^{-44}$ cm$^{2}$
is the neutrino cross section coefficient. On the other hand,
neutrino-electron scattering process in a strong magnetic field has
been discussed by many authors using various approaches (e.g.,
Bezchastnov \& Haensel 1996; Kuznetsov \& Mikheev 1997, 1999; Hardy
\& Thoma 2000; Mikheev \& Narynskaya 2000). Although the analytic
results are quite complicated and different based on various
approximated treatments, all of these works show that the cross
section $\sigma_{\nu_{i}e}$ is proportional to the field strength
$\propto eB_{m}$. For simplicity, we use the factor $2eB_{m}\hbar
c/(k_{B}T)^{2}$ to compare the importance of electron energies in
strong fields and the thermal energy, and consider the cross section
$\sigma_{\nu_{i}e}$ in strong magnetic fields as
\begin{equation}
\sigma_{\nu_{i}e}\simeq\sigma_{\nu_{i}e(B=0)}\times[1+2eB_{m}\hbar c/(k_{B}T)^{2}]
\end{equation}
with
\begin{equation}
\sigma_{\nu_{i}e(B=0)}=\frac{3\sigma_{0}}{8}\left(\frac{k_{B}T}{m_{e}c^{2}}\right)\left(\frac{\varepsilon_{\nu}}{m_{e}c^{2}}\right)
\left(1+\frac{\eta_{e}}{4}\right)[(C_{V}+C_{A})^{2}+(C_{V}-C_{A})^{2}/3].\label{cooling6}
\end{equation}

\section{Conservation Equations in Magnetized Disks and Field Topology}

\subsection{Conservation Equations}

In Section 2 we discussed the microphysics and thermodynamics
equations in strong magnetic fields, and compare them with the
equations without fields. Now we consider the basic conservation
magnetohydrodynamical (MHD) equations for a vertically-integrated
steady-state magnetized disk. We modify the hydrodynamical equations
without fields by adding the coupling of a large-scale magnetic
field. In this case we consider the accretion flow is still
constrained in the disk plane. The disk with an outflow structure
will be discussed in Section 5.1, and the funnel accretion process
in an extremely strong field in the stellar magnetosphere will be
discussed in Section 5.2. First of all, the mass continuity equation
in a vertically-integrated disk does not change
\begin{equation}
\dot{M}=-2\pi r\Sigma v_{r},\label{mass01}
\end{equation}
where $\Sigma=2\rho H$ is the disk surface density and $v_{r}$ is
the radial velocity of the accretion flow.

The vertically-integrating angular momentum conservation equation
with fields reads (e.g., Lovelace et al. 1987; Shadmehri \&
Khajenabi 2005)
\begin{equation}
\dot{M}\frac{dl}{dr}=-2\pi\frac{d}{dr}\left(r^{3}\Sigma\nu\frac{d\Omega}{dr}\right)-r^{2}B_{z}B_{\phi}|_{z=H}
+\frac{dH}{dr}(r^{2}B_{r}B_{\phi})|_{z=H}-\frac{d}{dr}(Hr^{2}\langle B_{r}B_{\phi}\rangle),
\end{equation}
where $\langle\ldots\rangle=\int_{-H}^{H}dz(\ldots)/2h$, $\nu=\alpha
c_{s}H$ is the kinematic viscosity coefficient in the disk with
$\alpha$ being the viscosity parameter. Here we still use the
$\alpha$-prescription\footnote{Actually, the half thickness of the
disk $H$ and the kinematic viscosity $\nu=\alpha c_{s}H$ will change
depending on the field structure. For example, the half thickness
$H$ will increase for large $B_{\phi}$ if $B_{r}\ll B_{\phi},B_{z}$
(Zhang \& Dai 2008b). Also, Shakura \& Sunyaev's scenario (1973)
$\nu=\alpha c_{s}^{2}/\Omega_{K}$ should be modified as $\nu=\alpha'
c_{s}^{2}/\Omega_{K}$ with $\alpha'$ being the function of magnetic
fields. In this paper we use the following treatment: calculate the
magnetic pressure $P_{\rm B}$ into the total pressure, since the
isothermal sound speed $c_{s}\propto P^{1/2}$, $c_{s}$ and the disk
half thickness can increase for strong fields. We still adopt the
classical relation $H=c_{s}/\Omega_{K}$ and $\nu=\alpha
c_{s}^{2}/\Omega_{K}$ for simplicity.}. The magnetic field can play
an important role in transferring the angular momentum. If the
radial field $B_{r}$ can be neglected (as will be shown in Section
3.2), we can integrate the angular momentum equation as
\begin{equation}
\frac{\dot{M}}{3\pi}f=\nu\Sigma-\frac{\dot{M}N_{B}}{3\pi
r^{2}\Omega},\label{angmom02}
\end{equation}
where
\begin{equation}
\dot{M}N_{B}(r)=-\int_{r}^{\infty}r^{2}B_{z}B_{\phi}|_{z=H}dr
\end{equation}
is the integrated magnetic torque, $f=1-l_{0}/(r^{2}\Omega)$ with
$l_{0}$ being the specific angular momentum constant. In the
standard assumption that the viscous torque is zero at the inner
boundary of the disk $r_{*}+b$ with $r_{*}$ being the stellar radius
and $b\ll r_{*}$, we can take $f=1-\sqrt{r_{*}/r}$ (Frank et al.
2002).

The local energy conservation equation in a magnetized disk is
\begin{equation}
Q_{\rm vis}^{+}+Q_{\rm Joule}^{+}=Q_{\rm adv}^{-}+Q_{\nu}^{-},\label{energy01}
\end{equation}
where $Q_{\rm vis}^{+}$ and $Q_{\rm adv}^{-}$ keep the same as in
the case of $B_{m}=0$ ((\ref{a12}) and (\ref{a13}) in Appendix A),
and the neutrino cooling rate $Q_{\nu}^{-}$ in strong magnetic
fields have been discussed in Section 2.2. In particular, the Joule
dissipation (or say the Joule heating) term $Q_{\rm Joule}^{+}$ is
\begin{eqnarray}
Q_{\rm Joule}^{+}&=&\frac{4\pi H}{c^{2}}\eta_{t}\langle J^{2}\rangle=\frac{H}{4\pi}\eta_{t}\langle(\nabla\times B_{m})^{2}\rangle\nonumber
\\&&\simeq \frac{H\eta_{t}}{4\pi}\left\{-\frac{2B_{r}|_{z=H}}{H}\frac{\partial B_{z}}{\partial r}|_{z=H}+\left(\frac{\partial B_{z}}{\partial r}\right)^{2}
+\frac{1}{r^{2}}\left[\frac{\partial(rB\phi)}{\partial
r}\right]^{2}\right\},\label{energy03}
\end{eqnarray}
where the magnetic diffusivity parameter $\eta_{t}$ is adopted as
$\eta_{t}\sim\nu\simeq\alpha c_{s}H$ (Lovelace et al. 1995;
Bisnovatyi-Kogan \& Lovelace 1998). In Section 6.1, we will discuss
the Joule dissipation term in more details.

Moreover, we need to add the charge conservation equation and the
chemical equilibrium equation, which are the same in paper II (or
see Appendix A (\ref{a07}) and (\ref{a14})).

\subsection{Field Topology}

The magnetic field of a neutron star or magnetar threading the disk
has the vertical component
\begin{equation}
B_{z}(r)=B_{0}\left(\frac{r_{*}}{r}\right)^{n}\label{t01}.
\end{equation}
The actual magnetic field in the vertical component near a magnetar
may be a mix of monopole ($n=2$) and dipole ($n=3$) fields (Thompson
et al. 2004). The dipole field drops faster than the monopole field
along the disk radius, thus the disk properties will be quite
different with the other cases. We focus on the dipolar form $n=3$
in this paper.

The differential rotation between the disk and the star will
generate a toroidal field (Wang et al. 1995). Following Lai (1998),
we consider two possible field configurations in a disk, i.e., the
stellar magnetic field threads the accretion disk in a closed
configuration (i.e., the classical configuration as in Ghosh \& Lamb
1979a), or the magnetic field becomes open (e.g., Lovelace et al.
1995). The generated azimuthal component of the magnetic field in a
steady-state disk is
\begin{equation}
B_{\phi}|_{z=H}=-\beta B_{z}\label{fieldstru01}
\end{equation}
for an open magnetic field in the disk, and
\begin{equation}
B_{\phi}|_{z=H}=-\beta
\left(\frac{\Omega-\Omega_{*}}{\Omega_{K}}\right)B_{z}\label{fieldstru02}
\end{equation}
for a closed magnetic field, where $\beta$ is a dimensionless
parameter, $\Omega_{*}$ and $\Omega_{K}$ are the stellar surface
angular velocity and Kepler velocity respectively. The radial
component $B_{r}$ generated in the disk reads
\begin{equation}
B_{r}|_{z=H}\propto\left(\frac{-v_{r}}{\Omega_{K}r}\right)B_{z}|_{z=H}
\end{equation}
In most cases, we consider $v_{r}\ll v_{K}=\Omega_{K}r$ and
$B_{r}|_{z=H}\ll B_{z}|_{z=H}$. Thus the field strength $B_{m}$ in
the disk can be obtained as
$B_{m}=\sqrt{B_{z}^{2}+B_{\phi}^{2}+B_{r}^{2}}\approx\sqrt{B_{z}^{2}+B_{\phi}^{2}}$.

According to Lovelace et al. (1995), if the angular velocities of
the star and disk differ substantially, the magnetic field lines
threading the star and the disk undergo a rapid inflation so that
the field becomes open. As a result, the outer part of the disk will
maintain an open field configuration, while the inner disk in the
magnetosphere still has closed lines. Since in the hyperaccreting
case the magnetosphere is very small except for an extremely strong
center field, in which the disk plane will be disrupted and
truncated in the magnetosphere (see discussion in Section 5), the
magnetized disk where still has a plane geometry would be more
favorable to maintain an open field. However, we still consider the
two cases as a broader consideration. If the magnetic field in the
disk is open, the magnetic torque in equation (\ref{angmom02})
becomes
\begin{equation}
\dot{M}N_{B}=\beta
B_{z}^{2}\frac{r^{3}}{2n-3},
\end{equation}
and the integrated angular momentum equation (\ref{angmom02}) can be
written as
\begin{equation}
\frac{GM\dot{M}}{3\pi}f+\frac{1}{3\pi}\frac{\beta}{s}\sqrt{GM}B_{z}^{2}\frac{r^{5/2}}{2n-3}=2\alpha\frac{P^{3/2}}{\rho^{1/2}}r^{3},\label{angmom04}
\end{equation}
where we take the ratio of the angular velocity of the disk flow and
the Keplerian velocity as a constant $s=\Omega/\Omega_{K}$. For the
case of a closed disk field, on the other hand, using
(\ref{fieldstru02}) we have
\begin{equation}
\dot{M}N_{B}=\beta
sB_{z}^{2}\frac{r^{3}}{2n-3}-\beta\Omega_{*}B_{z}^{2}(GM)^{-1/2}\frac{r^{9/2}}{2n-9/2}\label{angmom05}
\end{equation}
and the angular momentum equation (\ref{angmom02}) reads
\begin{equation}
\frac{GM\dot{M}}{3\pi}f+\frac{\sqrt{GM}}{3\pi}\frac{\beta}{2n-3}B_{z}^{2}r^{5/2}-\frac{2\beta}{3s}\frac{B_{z}^{2}r^{4}}{P_{r}}\frac{1}{2n-9/2}\label{angmom06}
=2\alpha\frac{P^{3/2}}{\rho^{1/2}}r^{3},
\end{equation}
where $P_{r}$ is the spin period of the magnetar.

Moreover, we obtain the energy conservation equation
(\ref{energy01}) in disk with the detailed field topology structure
as
\begin{equation}
\frac{3GM\dot{M}}{8\pi r^{3}}f+\frac{H\eta_{t}}{4\pi r^{2}}[n^{2}+(n-1)^{2}\beta^{2}]B_{z}^{2}=Q_{\rm adv}^{-}+Q_{\nu}^{-}\label{energy04}
\end{equation}
for an open magnetic field, and
\begin{equation}
\frac{3GM\dot{M}}{8\pi r^{3}}f+\frac{H\eta_{t}}{4\pi r^{2}}\left\{n^{2}+\left[s(n-1)-\frac{\Omega_{*}}{\Omega_{K}}\left(n-\frac{5}{2}\right)\right]^{2}\beta^{2}\right\}B_{z}^{2}=Q_{\rm adv}^{-}+Q_{\nu}^{-}\label{energy05}
\end{equation}
for a closed field.

We make a brief summary in the end of this section. We consider the
quantum effects of strong fields on the disk thermodynamics and
microphysics in Section 2.1, and the MHD conservation equation with
coupling of a large scale disk field in Section 2.2. The strong
magnetic field can change the electron distribution and energy
significantly, and also change the cross sections of the various
neutrino reactions. The large scale field can play an important role
in transferring angular momentum besides the viscous stress, and
heat the disk as well. In order to see the differences of disks with
and without field (or more exactly to say, with fields $\leq10^{14}$
G in the hyperaccreting disk) clearly, we list the basic equations
without fields in Append A for a comparison.

\section{Numerical Solutions of Magnetized Neutrino-Cooled Disks}

We adopt the $\alpha$ turbulent viscosity model of Shakura \&
Sunyaev (1973) and fix $\alpha=0.1$ in all of our calculations.
First we want to study the direct quantum effects of strong magnetic
fields on the disk, and the macrophysical field coupling in the MHD
conservation equations independently. Figure 4 and Figure 5 give the
independent results. In Figure 4, we keep the conservation equations
as the normal hydrodynamical equations without fields (i.e.,
equations (\ref{a08}), (\ref{a10}) to (\ref{a14}) in the Appendix),
and only adopt the equations of state (\ref{energy00}) to
(\ref{cooling6}) in Section 2 to see the quantum effects of a field.
On the other hand, in Figure 5, we keep the microphysical and
thermodynamical equations as in the case of no field (i.e.,
equations (\ref{a01}) to (\ref{a07}) in Appendix A and the equations
in Section 2 without fields), but only discuss the field effect on
angular momentum transfer (\ref{angmom02}) and energy heating
(\ref{energy01}). As discussed by Duncan \& Thompson (1992), in
principle the magnetic fields as strong as $3\times10^{17}(P/{\rm
ms})^{-1}$ G can be generated in magnetars as the differential
rotation smoothed by growing magnetic stresses. Klu\'{z}niak \&
Ruderman (1998, see also Ruderman et al. 2000) argued that the
energy stored in the differentially-rotating magnetars can be
extracted by the process of wounding up and amplification of
toroidal magnetic fields inside the star up to $\sim10^{17}$ G, then
the ultra strong field will be pushed to and through the surface by
buoyancy force. Based on these considerations, we adopt the magnetar
surface field up to $\sim10^{17}$ G, and discuss the field strength
from $10^{14}$ G to $10^{17}$ G with a dipolar extension in our
calculations. The steady state disk can be considered as the
transient hyperaccreting disk with a fixed time, or say, with a
fixed accretion rate. On the contrary, for a time-dependent disk
model (e.g., Janiuk et al. 2004, 2007 without fields), we have to
consider the evolution of the magnetar surface field, which is
beyond the purpose of this paper. Moreover, we do not consider other
quantum processes in strong fields such as the generation of strong
electric field and electron/positron pairs as $E\rightarrow
e^{-}+e^{+}+E$ and $\gamma+B\rightarrow e^{-}+e^{+}+B$ (Usov 1992).
These processes are important in the pulsar magnetosphere, but plays
a secondary role in the disk region outside the Alfv\'{e}n radius.
In Figure 4, a stronger magnetic field makes the disk be thinner,
cooler with lower temperature and neutrino luminosity. On the
contrary, Figure 5 shows that the disk will be denser with higher
pressure and neutrino luminosity, and becomes hotter at least in the
inner disk region near the magnetar, because the stronger field
plays a more significant role in transporting the disk angular
momentum and heating disk by Joule dissipation\footnote{In this
section, we extend our calculation to the inner disk region inside
the Alfv\'{e}n radius $r_{A}$. This extension is good for us to see
the disk properties of magnetized disks, although we need to use
another treatment to discuss the disk structure in the magnetosphere
with an extreme strong field.}. As a result, the quantum effects in
strong fields and the field coupling in MHD equations play two
opposite roles in changing the disk properties, one to decease
pressure, density and luminosity with increasing field strength, and
the other to increase them. These two competitive factors work
together to establish the actual structure of the disk.

In Figure 6 we consider the quantum and coupling effects together.
This figure is for the disk structure and neutrino luminosity with
accretion rate $\dot{M}=0.1M_{\odot}$ s$^{-1}$ and an open magnetic
field configuration in the disk. From equation (\ref{angmom04}) we
know that the ratio between the two parameters $\beta/s$ determines
the angular momentum transfer by magnetic field. As the result of
Figure 6, a higher ratio of $\beta/s$ leads to a higher density and
pressure in the entire disk, higher temperature and neutrino
luminosity in the inner region of the disk, as well as lower
electron fraction at the disk radius $\sim30$ km. Moreover,
electrons in the disk becomes more degeneracy around $\sim20-40$ km
for higher $\beta/s$, but quickly changes to be nondegeneracy at the
inner edge of the disk toward the magnetar surface.

Figure 7 gives the magnetized disk for various surface vertical
filed $B_{0}$ and open magnetic field configuration in the disk.
Stronger fields give higher density, pressure, temperature and
neutrino luminosity. This figure shows that the magnetic field
coupling in the MHD equations of disk flow is more important than
the quantum effects of strong fields in a microphysical scale. In
the case of $B_{0}=10^{17}$ G, the disk has no steady state solution
in the inner region of the disk where the flow should co-rotate with
the central star and further be channeled onto the magnetic pole
along the field lines. We will discuss the funneled flows later. On
the other hand, for $B_{0}=10^{14}$ G, we stop calculation where the
disk field is too weak for us to consider the quantum effect in
strong fields as discussed in Section 2. In this case, the disk
properties will be very similar to the case of $B_{0}=10^{14}$ G
without quantum effect even in the case of $B_{0}=0$, which has been
showed in Figure 5. The electron fraction decreases inward for
$B\leq10^{16}$ G, which is similar to the results without fields
(e.g., paper I, Fig. 7), but a field $\sim10^{17}$ G will change the
electron fraction distribution significantly. The distribution of
electron potential $\eta_{e}$ along the radius also changes a lot in
strong fields. Larger peak with more degeneracy electron state is
obtained by stronger fields, while the change of $\eta_{e}$ becomes
less obvious for relatively weaker fields.

Above we focus on the disk accretion rate of $\sim0.1M_{\odot}$
s$^{-1}$, which is typical for a magnetar hyperaccreting disk.
Moreover generally, we can study the magnetized disks with different
accretion rates. We discuss the disk properties with open disk field
and different accretion rate $\dot{M}=0.02, 0.1$ and 1 $M_{\odot}$
s$^{-1}$ in Figure 8. The values of density, pressure temperature
and neutrino luminosity become higher for higher accretion rate.
This result is similar to the case without field. Also, lower
electron fraction and more degeneracy electron state can be obtained
for higher accretion rate, except for the inner region where the
Joule dissipation becomes more significant than the viscous heating.
Therefore, the accretion rate plays a very similar role in changing
the disk properties to that without fields. However, we should point
out that, there is no steady-state solution for
$\dot{M}=0.02M_{\odot}$ s$^{-1}$ in the inner disk region
$B_{0}=10^{16}$ G. This is reasonable and similar to the case of
$\dot{M}=0.1M_{\odot}$ s$^{-1}$ and $B_{0}=10^{17}$ G, as such a
region is already inside the Alfv\'{e}n radius and is almost belong
to the magnetosphere. The extension calculation in this section
cannot reach this region.

Figure 9 shows the disk structure for a closed magnetic field in the
disk. We discuss it besides the open disk field configuration for
completeness. In this case, the disk properties not only depend on
the disk field structure ($\beta$) and angular velocity ($s$), but
also depend on the spin period of the central magnetar. A normal
pulsar has a spin period around 1 s $\sim$ 100 s, while the new born
pulsar as a candidate for GRB central engines usually has a much
shorter period with a timescale about tens of milliseconds or even
less. Here we consider the magnetar with a period $P_{r}=5,10,100$
ms. A rapidly rotating magnetar can act an extra torque on the disk.
However, we only consider the magnetic field torques from the
magnetar for simplicity. A shorter period of the central magnetar
decreases the disk density, pressure and electronic chemical
potential slightly, but increases the temperature at $20-40$ km and
the electron fraction at $r\geq20$ km (note that there is a peak of
$\eta_{e}$) at $r=20$ km for $B_{0}=10^{16}$ G. The effects of
period change on the disk are only obvious for the surface vertical
field $B_{0}\geq10^{16}$ G in Figure 9. However, since the neutrino
luminosity does not change significantly even for
$B_{0}\simeq10^{16}$ G, we cannot expect any obvious observational
events from the disk plane related to the effect of period on the
disks directly.

Figure 10 shows the $\dot{M}-\Sigma$ curves for a given disk radius
$r=40$ and 80 km. In order to see the stable performance clearly, we
extend our calculation to $\dot{M}=10M_{\odot}$ s$^{-1}$, although
such an accretion rate is impossible for an accreting magnetar
system. The condition for viscous stability is $d\dot{M}/d\Sigma>0$.
Therefore the disk is viscously stable for $B_{0}\leq10^{16}$ G, but
becomes unstable in the inner region $r\leq40$ km for
$B_{0}\sim10^{17}$ G with both open and closed disk fields until
$\dot{M}\geq1 M_{\odot}$ s$^{-1}$. However, because the Alfv\'{e}n
radius $r_{A}\gg40$ km for $\dot{M}\leq1M_{\odot}$ s$^{-1}$ and
$B_{0}\sim10^{17}$ G, such an entire instable region actually
interacts in the stellar magnetosphere. Therefore we can conclude
that the accretion flow in the disk plane can always be viscously
stable. This conclusion is consistent with that in the case of no
fields (Narayan et al. 2001; Di Matteo et al. 2002; Kawanaka \&
Mineshige 2007), but is somewhat different to that in Lei et al.
(2009), who showed that the disk is always unstable in the inner
disk region. The main difference is caused by different magnetic
structures between neutron star disks and their black hole
counterparts.

The black hole hyperaccreting disk without fields is thermally
stable, at least in the region where the nucleon gas pressure
dominates over the total pressure (Narayan et al. 2001; Di Matteo et
al. 2002, Kawanaka \& Mineshige 2007). However, a magnetic field in
the disk can make the disk be thermally unstable. We roughly
estimate it with an analytic method. The general condition for
thermal stability can be taken as (Narayan et al. 2001)
$(d\textrm{ln}Q^{+}/d\textrm{ln}H)|_{\Sigma}<(d\textrm{ln}Q^{-}/d\textrm{ln}H)|_{\Sigma}$.
The ratio of viscous and Joule heating is
\begin{equation}
\frac{Q^{+}_{\rm vis}}{Q^{+}_{\rm Joule}}\propto\frac{GM\dot{M}f}{rH\eta_{t}B_{z}^{2}}
\propto\frac{\nu\Sigma}{\eta_{t}H}\propto H^{-1},\label{stab01}
\end{equation}
As a result we have
\begin{equation}
\frac{d\,\textrm{ln}Q^{+}_{\rm vis}}{d\,\textrm{ln}H}|_{\Sigma}<\frac{d\,\textrm{ln}Q^{+}}{d\,\textrm{ln}H}|_{\Sigma}
<1+\frac{d\,\textrm{ln}Q^{+}_{\rm vis}}{d\,\textrm{ln}H}|_{\Sigma},\label{stab02}
\end{equation}
i.e., the value of $(d\textrm{ln}Q^{+}/d\textrm{ln}H)|_{\Sigma}$
increases by unity if $Q^{+}_{\rm Joul}$ dominates over $Q^{+}_{\rm
vis}$ in the magnetized disk. Now we calculate the ratio
$(d\textrm{ln}Q^{+}_{\rm vis}/d\textrm{ln}H)|_{\Sigma}$ in the
magnetized disk. Taking the open field configuration as an example.
The ratio of angular momentum transferred by magnetic torque and
viscous stress is
\begin{equation}
\frac{\dot{J}_{\rm m}}{\dot{J}_{\rm vis}}=\frac{\beta}{s(2n-3)}\frac{B_{z}^{2}}{\sqrt{GM}\dot{M}f}\frac{r_{*}^{6}}{r^{7/2}}
=12.3\beta s^{-1}M_{1.4}^{-1/2}M_{-1}^{-1}f^{-1}r_{*,6}^{6}r_{6}^{-7/2}B_{0,16}^{2}.\label{stab03}
\end{equation}
Furthermore
\begin{equation}
\frac{d\,\textrm{ln}Q^{+}_{\rm vis}}{d\,\textrm{ln}H}|_{\Sigma}=2\left(1+\frac{\dot{J}_{\rm m}}{\dot{J}_{\rm vis}}\right)
=2+24.6\beta s^{-1}M_{1.4}^{-1/2}M_{-1}^{-1}f^{-1}r_{*,6}^{6}r_{6}^{-7/2}B_{0,16}^{2}.\label{stab04}
\end{equation}
We can see that in the outer part of the disk, the ratio
(\ref{stab04}) is still $\simeq2$, but it can be significant
increased in the disk inner part where the magnetic field is
sufficiently high. If we take the radius $r\simeq r_{A}$ or
$r_{6}\simeq
2\dot{M}_{-1}^{-2/7}M_{1.4}^{-1/7}B_{0,16}^{4/7}r_{*,6}^{12/7}$,
equation (\ref{stab04}) becomes
\begin{equation}
\frac{d\,\textrm{ln}Q^{+}_{\rm vis}}{d\,\textrm{ln}H}|_{\Sigma}\simeq2+2.174\beta s^{-1}(1-0.707\dot{M}_{-1}^{1/7}B_{0,16}^{-2/7})^{-1}.\label{stab05}
\end{equation}
On the other hand, for the neutrino-cooled region, the neutrino
cooling rate under strong fields can be approximately taken as
$Q_{\nu}^{-}\propto\Sigma T^{4}\propto H^{8}$, or
$(d\textrm{ln}Q^{-}/d\textrm{ln}H)|_{\Sigma}\simeq8$. Based on the
above calculation, whether the gas dominated disk region is
thermally stable or not depends on the disk radius, magnetic field
strength, disk angular velocity and accretion rate. For typical
values $\beta\sim s, \dot{M}_{-1}=1, B_{0,16}=1$, and $r$ is
comparable to $r_{A}$, we have
$(d\textrm{ln}Q^{+}/d\textrm{ln}H)|_{\Sigma}\simeq10.4$, which shows
a thermally unstable performance. Furthermore, The radiation
dominated region of accretion disks are expected to be thermally
unstable in the $\alpha$-model without a large-scale field (Lightman
\& Eardley 1974; Shakura \& Sunyaev 1973; Narayan et al. 2001),
therefore such a region in the magnetized disk around a magnetar
will be always thermally unstable. On the other hand, Hirose, Krolik
\& Blaes (2009) showed the disk thermal stability based on the
3-dimensional radiation MHD simulations of a vertically stratified
shearing box, in which the turbulent magnetic field is generated by
magneto-rotational instability. The main reason of that difference
is based on different microphysics consideration. The stress is
assumed to be proportional to the radiation pressure in the
$\alpha$-prescription, while the stress fluctuations precede
pressure fluctuations, or say the magnetic energy fluctuations drive
pressure fluctuations in the MHD simulation. However, their MHD
simulation focuses on the accretion rate near a fraction of the
Eddington limit with normal radiation process rather than neutrino
cooling. In our paper we still use the $\alpha$-prescription with
the large-scale field. Thus we have a thermal unstable conclusion in
the radiation-dominated regions of the magnetized disks.

\section{Disk Outflows and Funnel Flows}

So far we study the structure and luminosity of the magnetized disks
with an open or closed field configuration constrained in the disk
plane. If we define the Alfv\'{e}n radius $r_{A}$ as the point at
which the magnetic pressure $P_{B}=B_{m}^{2}/8\pi$ equals the ram
pressure of the accretion material $\rho v_{r}^{2}$, and the
magnetospheric radius $r_{m}$ as the point where the magnetic
pressure is approximately equal to the total matter pressure
$P_{B}\simeq P_{\rm matter}+\rho v_{r}^{2}$, the disk plane will be
disrupted near the magnetosphere with $r\leq r_{A}$ as shown in
equation (1) and Figure 1, and the disk flow will be prevented from
accretion in the disk plane at $r_{m}$. Then almost the entire disk
will be funneled along the magnetar field lines onto only a small
fraction region of the stellar surface at $r<r_{m}$. The value of
$r_{m}$ depends on the details of disk-magnetar interaction. In the
normal case such as accretion onto magnetized neutron stars in X-ray
binaries, we have $P_{\rm matter}\ll\rho v_{r}^{2}$. Thus $r_{m}$
can be calculated as $P_{B}\sim\rho v_{r}^{2}$. It is estimated that
$r_{m}$ is close to or slightly less than the Alfv\'{e}n radius (Lai
1998). Let us see the situation in the magnetized hyperaccreting
disks. Differently we have $\rho v_{r}^{2}<P_{\rm matter}$ or
$v_{r}<c_{s}$ in the disk plane for the hyperaccreting case. Figure
11 gives the pressure ratio of magnetic stress and matter stress,
and the cooling efficiency along the radius. In this Figure we
confine the calculation in the disk plane without considering the
funnel effect. We find that the pressure ratio keeps $P_{\rm
B}/P_{\rm matter}<1$ in the disk plane, thus the accretion flow can
hardly be lifted by the field pressure above the plane. However,
since the accretion flow begins to co-rotate with the magnetar
surface inside the Alfv\'{e}n radius, the viscous heating and
angular transport which are generated by differential rotation in
the disk become ignorable. Also, the generated azimuthal component
of the disk field, which is proportional to the difference between
the disk and magnetar surface angular velocity, should also be
ignorable. As a result, the disk begins to cool and loss its angular
momentum. The matter pressure $P_{\rm matter}$ drops at the radius
inside the Alfv\'{e}n radius. In addition, as the magnetic pressure
increases as the accretion flow moving towards the magentar surface,
the ratio $P_{\rm B}/P_{\rm matter}$ keeps to increase and can be
greater than unity as well. Thus the magnetosphere (with its edge at
$r_{m}$) can actually form inside the Alfv\'{e}n radius, i.e., we
still have the magnetosphere region with $r_{*}<r_{m}<r_{A}$ in the
magnetized hyperaccreting disks.

Furthermore, the strong field in the Alfv\'{e}n radius would cause a
magnetically-driven outflow from the inner disk region; the
interaction between the disk and central star is favorable for
launching an X-type wind from the disk-magnetosphere boundary (Shu
et al. 1994), both of which also decrease the pressure of the
accretion matter and increase the ratio $P_{\rm B}/P_{\rm matter}$.
On the contrary, it is most likely that an thermally-driven outflow
can form in this outer region $r>r_{A}$. Generally speaking, the
thermal outflow can take away angular momentum and energy outside
$r_{A}$, make the direct flow accretion rate into the magnetosphere
and magnetar surface be low and give a possible energy source for a
supernova explosion associated with a GRB.

In Section 5.1, we discuss the disk outflow caused by thermal
heating. In Section 5.2 we discuss the funnel flow process due to
the co-rotation and magnetic pressure near the stellar
magnetosphere.

\subsection{Disk Outflows}

The thermal wind is induced by viscous heating (MacFadyen \& Woosley
1999), or neutrino heating (Metzger et al. 2008b), depending on the
dominated heating and cooling processes in the disk. In paper II, we
study the structure of the neutron star disk outflows around neutron
stars without fields. As shown by Narayan \& Yi (1994, 1995), if the
adiabatic index $\gamma< 3/2$ in an advection-dominated flow, the
Bernoulli constant of the flow is positive and a thermally driven
wind can be driven from the disk. In this case, the energy carried
by the outflow is expected to feed a supernova explosion, which is
associated with a GRB in the collapsar scenario (MacFayden \&
Woosley 1999; Kohri et al. 2005). This thermally-driven wind can
also exist in the magnetized disks beyond the Alfv\'{e}n radius,
where the matter pressure dominates over the magnetic pressure, and
heating energy generated by viscous and Joule dissipation is
advected inward along the disk radius. As showed in Figure 11 (right
panels), the neutrino cooling efficiency increases with increasing
the strength of a magnetic field for $\dot{M}\sim0.1 M_{\odot}$
s$^{-1}$, and the entire disk beyond the Alfv\'{e}n radius becomes
neutrino-dominated for $\dot{M}\sim 1.0M_{\odot}$ s$^{-1}$.
Therefore the outflow-driven process should be significant below
$\dot{M}\sim 1.0\dot{M}_{\odot}$ s$^{-1}$. In this section, we adopt
the similar treatment as in Kohri et al. (2005) and paper II for
simplicity, i.e., we only consider the viscously induced outflows
and assume the accretion rate varies as a power law in radius for
the case $\dot{M}< 1.0M_{\odot}$ s$^{-1}$ as
\begin{equation}
\dot{M}=\dot{M}_{\rm out}\left(\frac{r}{r_{\rm out}}\right)^{\xi},\label{wind01}
\end{equation}
where $\dot{M}_{\rm out}$ is the initial accretion rate at the outer
edge of the disk, and $\xi$ is the outflow index in the disk.
Stronger outflows have a lower value of $\xi$. Using the outflow
structure (\ref{wind01}), the angular momentum equation
(\ref{angmom02}) is modified as
\begin{equation}
\frac{1}{1+2\xi}\frac{\dot{M}}{3\pi}f=\nu\Sigma-\frac{\dot{M}N_{B}}{3\pi
r^{2}\Omega}.\label{wind02}
\end{equation}
Here we have assumed that the outflow has the same angular velocity
as the accretion flow outside the Alfv\'{e}n radius and takes away
angular momentum from the disk.

Figure 12 gives the disk structure with a viscous and Joule heating
induced thermally-driven outflow for various outflow index $\xi=0.2,
0.6, 0.9$, different magnetar surface fields $B_{0}=10^{16},
10^{17}$ G, and a fixed accretion rate $\dot{M}=0.5M_{\odot}$
s$^{-1}$ in the radius $r=124$ km (i.e., 30$R_{sh}$ of a 1.4
$M_{\odot}$ star). Similar to the results in paper II, the density,
pressure and neutrino luminosity decrease with increasing the
outflow strength. On the other hand, the ratio of $P_{\rm B}/P_{\rm
matter}$ and the thickness of the disk become larger for stronger
outflows. Therefore a disk with stronger thermally-driven outflow
outside the Alfv\'{e}n radius carries more energy away into the
stellar envelope at a fixed radius. However, the total outflow
energy strongly depends on the size of the region outside the
Alfv\'{e}n radius, where generates viscous and Joule heating energy
and induces outflows. A strong central stellar field makes the
Alfv\'{e}n radius be larger and decrease the total heating and the
potential outflow injection energy. Table 1 gives our estimate of
the maximum energy carried by the outflow, which can be calculated
as the difference between heating energy with and without filed,
\begin{eqnarray}
\dot{E}_{\rm out,max}&\sim&\int_{r_{A}}^{r_{\rm out}}Q^{+}_{\rm heat}(\xi=0)2\pi rdr-\int_{r_{A}}^{r_{\rm out}}Q^{+}_{\rm heat}(\xi)2\pi rdr\nonumber\\
&&=\frac{3GM\dot{M}}{4}\left\{\frac{1}{3r_{A}}-\frac{1}{r_{\rm out}}\left[1-\frac{2}{3}\left(\frac{r_{A}}{r_{\rm out}}\right)^{1/2}\right]\right\}\nonumber\\
&&-\int_{r_{A}}^{r_{\rm out}}\frac{9}{8}\nu\Sigma\frac{GM}{r^{3}}2\pi rdr+[\dot{E}_{\rm joul}(\xi=0)-\dot{E}_{\rm joul}(\xi)]\label{wind03}
\end{eqnarray}
The maximum energy injection rate decreases significantly for
stronger stellar fields, which make the magnetically-driven wind and
funnel effect be more important than the thermal outflow for strong
fields.

Besides the thermally-driven outflow in the disk outer region, a
strong field inside the Alfv\'{e}n radius causes the accretion flow
to co-rotate with the stellar field, and is possible to launch MHD
wind along the field lines (Metzger et al. 2008a). Moreover, it is
also clear that an X-type field configuration near the central star
is favorable for launching magnetically-driven wind from the
disk-magnetosphere boundary (Shu et al. 1994). As showed by
simulations (e.g., Romanova et al. 2008), the properties of these
MHD outflows depend on many factors such as the field structure, the
accretion rate, the related viscosity and magnetic diffusivity, etc.
The magnetically-driven winds in the disk inner region will decrease
the rate of accretion onto the magnetar surface. Also, the
magnetically-driven winds can provide energy to a supernova
explosion together with the thermal outflow. But this issue is
beyond the purpose of this paper.

\subsection{Funnel Flows from Disks to Magnetars}

When the magnetic pressure becomes equal or larger than the matter
pressure inside the Alfv\'{e}n radius $P_{\rm B}\geq P_{\rm
matter}$, the funnel accretion process becomes significant. For the
case of hyperaccreting disks, this process can only be important for
an extremely strong field, which depends on the accretion rate. For
simplicity, we consider the funnel process to take in the region
between the magnetosphere and the Alfv\'{e}n radius, and the
accretion in the disk plane will be truncated at the radius $r_{m}$.
The scenario of funneling process is similar to many previous work
(e.g. Lovelace et al. 1995. Fig 2, Fig 3; Koldoba et al. 2002, Fig
1; Frank et al. 2002, Fig 6.4). Different to the disk with
cylindrical coordinates, in this section we use the spherical
coordinates ($r_{l},\theta,\phi$) with origin at the stellar center
to describe the funneled flow, where $\theta$ is the angle between
the axis of the disk plane and a given polar radius $r$, $\phi$ is
the angle in the disk plane. Thus the equation for the dipole field
geometry line is (Frank et al. 2002)
\begin{equation}
r_{l}=r_{d}\frac{\textrm{sin}^{2}\theta}{\textrm{sin}^{2}\Theta},\label{f01}
\end{equation}
where $r_{d}$ is the radius at which the magnetic field line passes
through the disk, and $\Theta$ is the angle between the magnetic
pole and disk axis. We can approximately take $r_{m}<r_{d}<r_{A}$.

The magnetic field along the field line reads
\begin{equation}
B_{p}(r_{l})=B_{0}\left(\frac{r_{*}}{r_{l}}\right)^{3}(4-3\textrm{sin}^{2}\theta)=B_{0}
\left(\frac{r_{*}}{r_{l}}\right)^{3}\left(4-\frac{3r_{l}}{r_{d}}\textrm{sin}^{2}\Theta\right)^{1/2}.\label{f02}
\end{equation}
Note that the dipole field is somewhat different to the vertical
component $B_{z}(r)=B_{0}(r_{*}/r)^{3}$ in Section 3, in which $r$
is the radius of the disk in the disk plane. The mass conversation
along the magnetic flux line requires the poloidal part ($r,\theta$)
of the velocity $v_{p}$ and the density $\rho$ satisfies
\begin{equation}
\frac{4\pi v_{p}\rho}{B_{p}}=K,\label{f03}
\end{equation}
where $K$ is the constant along the flux line. Near the magnetar
surface, the accretion rate satisfies
\begin{equation}
4\pi r_{*}^{2}v_{*}\rho_{*}\left(\frac{\Delta\Omega}{2\pi}\right)=\dot{M},\label{f04}
\end{equation}
where $\Delta\Omega$ is the opening solid angle of the funnel flows,
$v_{*}$ and $\rho_{*}$ is the average velocity and density in this
angle $\Delta\Omega$ onto the stellar surface. We use the $2\pi$ as
the total solid angle because the funnel flow can be accreted
towards two poles. The difference between the final accretion rate
$\dot{M}$ onto the magnetar surface and the initial accretion rate
$\dot{M}_{\rm out}$ in the disk strongly depends on the strength of
magnetically and thermally driven winds. Different to the case of
Bondi accretion onto a magnetized compact star, which form an
accretion column to cover the magnetic pole, the funnel accretion
from a half-thickness disk forms a ring-like belt at the latitude of
$\theta$ between $\textrm{arcsin}\sqrt{r_{*}/r_{A}}$  and
$\textrm{arcsin}\sqrt{r_{*}/r_{m}}$ to cover a part of the stellar
surface. Recent simulations show that the geometry shape and
physical properties of this belt (or say the ``hot spot") are very
complicated, depending on the detailed field topology and the
structure of the funnel flows (Long et al. 2008, or see Romanova et
al. 2008 for a review). In this paper we take this ring-like belt as
axisymmetric around the magnetic pole axis for simplicity. Appendix
B gives our estimate of the ratio of $\Delta\Omega/2\pi$ in the case
of hyperaccreting and neutrino-cooled disks. Combining (\ref{f02}),
(\ref{f03}) and (\ref{f04}), we have
\begin{equation}
\rho v=\frac{\dot{M}(2\pi/\Delta\Omega)}{4\pi r_{*}^{2}}\left(\frac{B_{p}}{B_{p*}}\right).\label{f06}
\end{equation}
with $B_{p*}$ being the strength of the dipole field on the stellar
surface.

In the normal cases, the infalling funneled flow from the disk plane
onto a magnetized compact star usually will go through a strong
shock before reaching the stellar surface (e.g., Ferrari et al.
1985; Ryu et al. 1996; Li et al. 1996; Frank et al. 2002). However,
such a shock is probably unlikely to develop in the hyperaccreting
disks. Let us discuss it in more details. The Mach number
$\mathscr{M}$ in the funnel flow along the magnetic field pole can
be calculated as
\begin{equation}
\mathscr{M}^{2}=\frac{v^{2}}{a^{2}}=\frac{(v\rho)^{2}}{\gamma\rho P_{\rm matter}}
=\left(\frac{\dot{M}}{4\pi r_{*}^{2}}\right)^{2}\left(\frac{2\pi}{\Delta\Omega}\right)^{2}\left(\frac{B_{p}}{B_{p*}}\right)^{2}
\frac{1}{\gamma\rho P_{\rm matter}}\label{f07}
\end{equation}
with $\gamma$ being the adiabatic index of the disk matter. Thus
\begin{eqnarray}
\mathscr{M}^{2}&\sim&8\pi\left(\frac{\dot{M}}{4\pi r_{*}^{2}}\right)^{2}\left(\frac{2\pi}{\Delta\Omega}\right)^{2}
\left(\frac{B_{p}}{B_{p*}}\right)^{2}\left(\frac{P_{\rm B}}{P_{\rm matter}}\right)\frac{1}{\gamma B_{p}^{2}\rho}\nonumber\\
&&=6.30\times10^{-5}\gamma^{-1}\rho_{12}^{-1}\dot{M}_{-1}^{2}r_{*,6}^{-4}B_{p*,16}^{-2}(2\pi/\Delta\Omega)^{2}(P_{\rm B}/P_{\rm matter}),\label{f08}
\end{eqnarray}
where $\rho_{12}=\rho/10^{12}$ g cm$^{-3}$ and
$B_{p*,16}=B_{p*}/10^{16}$ G. For a typical hyperaccreting disk with
a stellar field $\sim10^{16}$ G and the accretion rate
$\dot{M}=0.1M_{\odot}$ s$^{-1}$, we have $P_{\rm B}\sim P_{\rm
matter}$, $\rho\sim10^{12}$ g cm$^{-3}$, and
$2\pi/\Delta\Omega\sim10^{2}$ (from Appendix B), therefore the Mach
number is $\mathscr{M}<1$, which shows there is no shock wave
existing in the funnel flow. A strong magnetically-driven wind will
make the density and pressure in the funnel flow decrease, but the
accretion rate $\dot{M}$ onto the star will also decrease. Although
$2\pi/\Delta\Omega$ can reach to $10^{3}$ and the factor $P_{\rm
B}/P_{\rm matter}$ increases for a strong field $\sim10^{17}$ G with
$\dot{M}=0.1M_{\odot}$ s$^{-1}$, but the decreasing factor
$\rho_{12}^{-1}B_{p*,16}^{-2}$ makes the Mach number hardly exceed
unity.

In the funnel channel, the flow is accelerated via the stellar
gravitational force. The gravitational binding energy is converted
to the kinematic energy of the flow, then the kinematic energy will
be converted to the heating energy near the magnetar surface, which
is cooled via thermal neutrino emission. The total energy release
rate in the funnel process can be approximately taken as
\begin{equation}
\dot{E}_{\rm funnel}\sim\frac{GM\dot{M}}{4}\left(\frac{1}{r_{*}}-\frac{1}{r_{A}}\right),
\end{equation}
and the energy equation in the magnetar surface boundary is
\begin{equation}
\frac{7}{8}\frac{\sigma_{B}T^{4}}{\tau_{\nu}}\cdot S=\frac{GM\dot{M}}{4r_{*}}\epsilon,
\end{equation}
where the parameter $\epsilon$ is introduced to show the combined
efficiency of acceleration and cooling. The accretion rate here is
adopted as the final accretion rate onto the magnetar. The area of
the ``hot spot'' is $S=\Delta\Omega r_{*}^{2}$. Here we consider the
new born magnetar with a lifetime $\gg100$ ms (Dessart et al. 2009),
then the temperature of the ``hot spot'' can be significantly higher
then the other region of the magnetar surface. The temperature in
the ``hot spot'' can be calculated as
\begin{equation}
T=7.4\times10^{10}(2\pi M_{1.4}\dot{M}_{-1}\epsilon/\Delta\Omega)^{1/4}r_{*}^{-3/4}\quad\textrm{K},
\end{equation}
where we take $\epsilon$ as 0.5. Since the temperature is
insensitive to the neutrino optical depth as $\tau_{\nu}^{1/4}$, we
take $\tau_{\nu}\sim1$ in our calculation. The temperature in this
surface region increases slightly with increasing the ratio of
$2\pi/\Delta\Omega$: $T=1.3\times10^{11}$ K ($10$ MeV) for
$2\pi/\Delta\Omega=10$, or $T=2.3\times10^{11}$ K ($20$ MeV) for
$2\pi/\Delta\Omega=100$.

The neutrino pair annihilation process
$\nu_{i}+\bar{\nu}_{i}\rightarrow e^{-}+e^{+}$ is the most important
mechanism to provide the energy for a relativistic jet formed from
the neutron star disk with weak stellar magnetic field. According to
paper II, the neutron star disk with a hot stellar surface layer
could increase the neutrino annihilation luminosity by about one
order of magnitude higher compared with the black hole disk. If the
stellar field is strong up to $\geq10^{15}$ G, as discussed in
Section 3, the disk with strong magnetic field will have a higher
density, pressure and neutrino luminosity, therefore the
annihilation rate above the disk plane will also be higher than that
emitted from a disk with the same radius range without field or with
weak field. This conclusion is similar to that in Lei et al. (2009,
their Fig. 2). However, the strong field truncate the disk plane
accreting in the inner region and will decrease the total
annihilation luminosity from the disk. In this case, the
annihilation mechanism will be more significant from the magnetar
surface where accretes the funnel flow. Similar to the case without
field, the total neutrino annihilation rate is contributed by three
components: the annihilation between neutrinos both from the disk,
both from the stellar surface, and one from the disk and the other
from the stellar surface respectively. If the stellar surface is
bright to neutrino emission, then the stellar surface will
contribute to the main annihilation luminosity in the accreting
system. This is why the neutron star accretion system will have a
brighter annihilation luminosity than its black hole counterpart. In
the funnel process, on the other hand, the stellar surface area
where has a solid angle $\Delta\Omega\ll2\pi$ and higher latitude
$\theta_{*}\sim \textrm{arcsin} \sqrt{r_{*}/r_{A}}$ to emit thermal
neutrino will be more geometrical concentrative and hotter than that
without field, so the annihilation rate will be more efficient
(Birkl et al. 2007). Figure 13 shows the neutrino annihilation rate
as a function of height along the magnetic poles with different
latitudes of the emitting ``hot spot" area. We adopt the formula of
neutrino annihilation as in Popham et al. (1999) and Rosswog et al.
(2003). The different point here is that the neutrinosphere is a
ring-like belt around the magnetar surface, but not a plane disk.
The neutrino annihilation rate $l_{\nu\bar{\nu}}$ from a higher
latitude ring-like belt area is larger near the stellar magnetic
poles, but becomes smaller far from the poles. That is because the
shorter distance of a given point along the pole to a higher
latitude emission area plays a key role in increasing the
annihilation efficiency. On the other hand, the less average value
of the angle or larger $\textrm{cos}\theta_{kk'}$ between neutrino
pairs from lower latitudes plays a more important role in increasing
the neutrino annihilation efficiency at a large height $\geq20$ km.
Whatever, the integrated neutrino annihilation energy along the pole
region $\int_{r_{*}}^{\infty}l_{\nu\bar{\nu}}dz$ is much higher from
the higher latitude ring-like belt than that from lower latitudes.
Therefore, funnel accretion can accumulate more powerful
annihilation luminosity (see the caption of Figure 13).

In the black hole disks with strong fields, the MHD effects such as
the Blandford-Znajek mechanism or magnetic instabilities (MRI) leads
to formation of a magnetically-dominated jet along the magnetic
poles. Such a magnetically-dominated jet will also form from the
magnetar in a few seconds of its formation, and the magnetar
rotation energy to feed the jet. Therefore in a hyperaccreting
magnetar system the neutrino annihilation mechanism and the MHD
mechanism will work together to form a jet, and make the jet be more
powerful for a GRB explosion. We will discuss this issue again in
Section 6.

Finally we list the properties of the hyperaccreting magnetar system
in Table \ref{tab04}, taking the accretion rate
$\dot{M}=0.1M_{\odot}$ s$^{-1}$ near the Alfv\'{e}n radius $r_{A}$
or the stellar surface $r_{*}$ (depends on $r_{A}>r_{*}$ or
$r_{A}<r_{*}$) as an example.

\section{Discussions}

\subsection{Joule Dissipation}

In Section 3 we discuss two competitive sets of effects to affect
the disk structure and neutrino emission, i.e., the microphysical
quantum effects and the macrophysics magnetic field coupling in the
disk MHD equations. The quantum effects decrease the pressure,
density and luminosity with increasing the strength of field, but
magnetic coupling makes these quantities be higher. However, in most
cases, the magnetic field coupling is more important than the
quantum effects in a microphysical scale. We discuss this conclusion
in details in this section. In fact, in order to see the quantum
effects clearly in Figure 4, we adopt a unified field (i.e, without
a dipolar form) in the disk, and find that this microphysical effect
becomes more significant in the disk region at relatively larger
radius. However, such a situation is unreal if we adopt the dipolar
field from the stellar surface, thus the microphysical quantum
effects will be less important as showed in Figure 4. On the other
hand, as showed in the angular momentum equations (\ref{angmom04}),
(\ref{angmom06}) and the local energy equation (\ref{energy04}),
(\ref{energy05}), stronger magnetic fields will be more important to
transfer angular momentum in the disk, as well as to heat the disk
by Joule dissipation. Therefore, it is not difficult to understand
why the disk with a stronger field will be hotter, denser with
higher pressure and brighter neutrino luminosity.

In fact, the parameter $\beta$ which gives the relation between the
vertical and azimuthal components of the disk magnetic field as in
equation (\ref{fieldstru02}, i.e., the closed field configuration)
can be calculated using the Ohm's law
$\textbf{j}=\sigma_{m}(\textbf{E}+\textbf{v}\times \textbf{B})/c$
and Ampere's law $\nabla\times \textbf{B}=4\pi \textbf{j}/c$. We
have (Lee 1999)
\begin{equation}
\beta=\frac{4\pi\sigma_{m}}{c^{2}}Hr\Omega_{K}=\frac{Hr\Omega_{K}}{\eta_{m}},\label{dis01}
\end{equation}
where $\sigma_{m}$ and $\eta_{m}$ are the microscopic electric
conductivity and magnetic diffusivity. However, as argued by some
authors (e.g, Bisnovatyi-Kogan \& Ruzmaikin 1976; Lovelace et al.
1995; Bisnovatyi-Kogan \& Lovelace 1997), since the accretion flow
are in general turbulent, the microscopic $\eta_{m}$ should be
replaced by a turbulent transport parameter $\eta_{t}$, which can be
considered as being comparable to the turbulent $\alpha$ viscosity
$\eta_{t}\sim\nu\sim\alpha c_{s}H$. In this case, equation
(\ref{dis01}) gives the relation $\beta\sim v_{K}\alpha^{-1}
c_{s}^{-1}$. However, this new relation has its limitations. As
discussed by Lovelace et al. (1995), the twist $|B_{\phi}/B_{z}|$
can never be much larger than unity. Therefore, in this paper, we
take $\beta$ as a parameter rather than an obtained variable.
Similar to Lai (1998), we consider that $\beta\sim1$ as the maximum
twist of the original magnetic field from the central stellar
surface. On the other hand, we take the magnetic diffusivity
$\eta_{t}$ in the local energy equation to be the order of the
turbulent viscosity. As a result the magnetic diffusivity can be
smaller than the classical diffusivity showed by Ghosh \& Lamb
(1979a, 1979b; see Lovelace et al. 1995 for further discussion).

The Joule heating plays a key role in affecting the neutrino
luminosity from the hyperaccreting disks with small radius, because
the ratio of the viscous heating and Joule dissipation will decrease
with small radius, and the Joule dissipation quickly dominates over
the viscosity as the main heating source in the disk. Of course all
of these results only occur in the region outside the Alfv\'{e}n
radius $r_{A}$, although we usually give our calculations extended
to the inner region. Figure 14 shows the neutrino cooling luminosity
without Joule dissipation. Compared with the neutrino luminosity
from the disk with Joule dissipation in Figures 5, 7, and 9, the
main difference is that the neutrino luminosity drops quickly in
relatively strong fields at small radius without fields, except for
the case that the disk has relatively low accretion rate and is
advection-dominated. This result is consistent with the above
discussion that Joule dissipation plays a key role in heating the
disk inner region near the Alfv\'{e}n radius.

\subsection{Application to GRBs}

As showed in Section 5.2, the funnel accretion makes the ring-like
belt of ``hot spot" be more concentrative with higher latitude than
the equatorial accretion without fields or with moderate fields.
Also, the neutrino luminosity outside the Alfv\'{e}n radius is also
brighter for stronger stellar fields. As a result, the neutrino
annihilation efficiency will be significantly increased in the
hyperaccreting disks around magnetars compared to the normal neutron
star disks, and definitely more efficient than the black hole disks.
However, a higher concentration neutrino emission region means
higher temperature on the magnetar surface with more massive
neutrino-driven winds (Qian \& Woosley 1996; Dessart et al. 2009),
which will make the jet be heavily baryon-loading. The problem is
that, whether the more powerful annihilation and massive mass loss
rate driven by neutrino absorption can work together to produce a
relativistic jet required for a GRB explosion? This problem is
similar to that in the neutron star disks without fields, which has
been discussed in paper II. We consider that a GRB-related jet forms
along the stellar poles (i.e., the disk axis if there is no field),
and only the wind materials along this axis can feed the jet and
affect the bulk Lorentz factor of the jet. The wind ejected in a
off-axis direction and that evaporated from the disk will not affect
the jet, but only affect the potential supernova explosion
associated with the GRB. If there is not disk around the central
compact star, the neutrino-driven wind from the star surface can be
reasonably approximated as a quasi-steady spherical outflow (Qian \&
Woosley 1996). However, the hyperaccreting accretion disk around the
star can change the distribution of the neutrino-driven wind
significantly. The entire stellar surface will be very hot and
inject a heavily mass-loading wind along the axis after its
formation in a timescale of $\sim100$ ms, and then to cool down with
the power law $\dot{M}_{\rm wind}\propto t^{-5/3}$. In most cases,
the surface ring-like belt region where directly accretes the disk
flow will be the hottest region in the disk for the time $\gg100$
ms. However, most parts of the wind from this ring-like belt should
be off-axis. The polar region that directly drives a wind along the
axis is cooler than the equatorial ring-like belt. Even though, we
use the temperature of the hot ring-like belt to estimate the
maximum strength of the neutrino-driven wind along the poles, and
calculate the bulk Lorentz factor of the wind along the pole with
field $\leq10^{15}$ G in paper II. A moderately or ultra
relativistic jet can be produced in the hyperaccreting neutron star
system with sufficient high disk accretion rate and bright boundary
emission.

However, the situation in a strong field will be different in some
aspects. First of all, the neutrino-to-nucleon absorption reaction
rate, which is dominated by
\begin{equation}
\nu_{e}+n\rightleftharpoons p+e^{-}
\end{equation}
\begin{equation}
\bar{\nu}_{e}+p\rightleftharpoons n+e^{+}
\end{equation}
will be significantly reduced in strong fields $B_{0}\geq10^{16}$ G
(Duan \& Qian 2004, 2005) because of the quantum effects. As a
result, the mass loss rate from the stellar surface which is
proportional to the absorption rates will decrease as well.
Moreover, we have to note that part of the newly generated
$e^{-}e^{+}$ particles in the neutrino absorption reaction will move
along the closed field lines from the stellar surface rather than be
injected far away from the central star. There are some open field
lines in the surface region near the stellar magnetic poles
(Lovelace et al. 1995), but many of them are more likely to induce
winds off-axis rather than the neutrino-driven wind. As a result,
the total mass loss rate $\dot{M}_{\rm polar}$ along the magnetic
polar region will be only a fraction of the total mass loss rate
$M_{\rm wind}$. On the other hand, the field structure around the
stellar surface will also affect the neutrino annihilation process,
because part of the generated $e^{-}e^{+}$ pairs in the reaction
$\nu_{i}+\bar{\nu}_{i}\rightarrow e^{-}+e^{+}$ will also move along
the closed field lines. Therefore, part of the neutrino annihilation
energy cannot be loaded into the jet which propagates along the
magnetic poles. The precise fraction of the total annihilation
energy feeding the polar jet depends on the energy-momentum
distribution of the $e^{-}e^{+}$ pairs above the central star and
the entire disk, as well as the ratio of $e^{-}e^{+}$ plasma
pressure to the magnetic pressure. The reliable estimate of these
considerations requires further MHD simulations. However, since the
annihilation process happens in the extended space above the
accreting system, while the neutrino absorption reaction mainly
occurs around the stellar surface with a sufficient high density of
nucleons, the strong magnetic fields will play more significant role
in changing the properties of the neutrino-driven wind rather than
the annihilation process. Therefore the mass loaded in the jet will
decrease more significantly than that of the annihilation energy.
Based on these above considerations, that is, the neutrino
annihilation efficiency is larger for a magnetized disk, such an
efficiency can even higher for the funnel accretion process, and the
mass loss along the magnetic poles $\dot{M}_{\rm wind}$ decreases in
strong fields, we can conclude that the jet along the magnetic poles
can be accelerated to a larger bulk Lorentz factor compared with its
neutron star counterparts without fields or with weak fields. In
addition, if the magnetic pole and the disk axis do not overlap
($\Theta\neq90^{\circ}$), the direction of the relativistic jet and
the disk rotation axis will also not overlap. This is an interesting
topic, because if the annihilation rate from the disk is much less
than that from the star, the jet will precess along the disk axis
with a period of the magnetar.

Furthermore, another significant difference between a magnetar and a
non-magnetized neutron star in the hyperaccreting systems is that, a
thermally-dominated neutrino-driven wind from the stellar surface
may switch to be magnetically-dominated instead after seconds of the
compact star formation (Thompson et al. 2004; Metzger et al. 2007).
The condition of the magnetic wind can be parameterized using the
term of magnetization
\begin{equation}
\sigma=\frac{\Phi_{B}^{2}\Omega_{*}^{2}}{\dot{M}_{\rm wind}c^{3}}
\end{equation}
where $\Phi_{B}$ is the magnetic flux. The parameter $\sigma$, which
strongly depends on the rotational period of the central star, is
the maximum Lorentz factor the wind can achieve if all magnetic
energy is converted into kinetic energy, either via magnetic
reconnection (Spruit et al. 2001) or collimation by the interaction
between the wind and the the stellar envelope (Bucciantini et al.
2006, 2007, 2009; Tchekhovskoy et al. 2008, 2009). The critical
period for $\sigma\geq1$, i.e., the wind to be
magnetically-dominated and relativistic is
\begin{equation}
P_{rc}\leq85B_{0,16}r_{*,6}^{2}\dot{M}_{\rm wind, -5}^{-1/2} \textrm{ms}.\label{grb1}
\end{equation}
The critical period $P_{rc}$ increases with increasing the stellar
field, and decreases with increasing the mass loss rate $\dot{M}$.
The properties of the magnetically-dominated wind will be different
with different ranges of $\sigma$ for the stellar period $P_{r}\leq
P_{rc}$. The wind is collimated along the magnetic poles for
$\sigma<5$, but will be distributed around the direction at low
latitudes and the equatorial plane for $\sigma>30$. For example,
Bucciantini et al. (2006, 2007, 2009) showed that, the
wind-stellar-envelope interaction may provide a viable mechanism for
collimating the jet along the polar region. Their works are based on
the consideration that a cavity with the radius $>10^{8}$ cm has
been evacuated by the outgoing supernova shock before the compact
star formation. Since the wind with high $\sigma$ is ejected around
the equatorial plane initially, whether or not the hyperaccreting
disk with a small size $10^{7}-10^{8}$ cm can help or prevent the
jet collimation needs to be further studied.

It is said that the energy of the magnetically-driven wind is
extracted by the central star spin-down process. However, in the
magnetar-disk system, if the magnetically-driven wind can be
actually collimated along the magnetic polar region, then the
stellar spin-down mechanism and neutrino annihilation from the ``hot
spot" and the entire disk can work together to provide energy of the
polar jet and accelerate the jet to an ultra relativistic speed. In
this general case, the bulk Lorentz factor can be estimated as
\begin{equation}
\Gamma=\frac{\dot{E}}{\dot{M}_{\rm polar}c^{2}},
\end{equation}
where the energy feed rate is the summation of neutrino annihilation
and the magnetic energy from rotational extraction,
\begin{eqnarray}
\dot{E}&=&\dot{E}_{\rm anni}+\dot{E}_{\rm mag}\nonumber\\
&&=L_{\nu\bar{\nu}}f_{k}+\dot{E}_{\rm rot}f_{\rm rot}=L_{\nu}(f_{\nu\bar{\nu}}f_{k})+\dot{E}_{\rm rot}f_{\rm rot}\nonumber\\
&&=\frac{GM\dot{M}}{4r_{*}}(\epsilon f_{\nu\bar{\nu}}f_{k})+\frac{2}{5}Mr_{*}^{2}\Omega\dot{\Omega}f_{\rm rot},
\end{eqnarray}
with $f_{\nu\bar{\nu}}$, $f_{k}$, $f_{\rm rot}$ being the neutrino
annihilation efficiency $L_{\nu\bar{\nu}}/L_{\nu}$, the fraction of
the deposited annihilation energy to provide the kinetic energy of
the stellar wind, and the ratio of magnetic to total spin-down
energy respectively. $\dot{M}_{\rm polar}$ as a fraction of
$\dot{M}_{\rm wind}$ is the wind mass loss rate along the polar
region (see paper II for more discussion about the mass loss rate).
The ratio of $\dot{E}_{\rm anni}$ to $\dot{E}_{\rm mag}$ is
\begin{equation}
\frac{\dot{E}_{\rm anni}}{\dot{E}_{\rm mag}}=2.10\times10^{5}\dot{M}_{-1}P_{r}^{2}\tau_{J}r_{*,6}^{3}\left(\frac{\epsilon f_{\nu\bar{\nu}}f_{k}}{f_{\rm rot}}\right),
\end{equation}
where $\tau_{J}$ being the typical spin-down timescale. Based on the
analysis in this paper, we take $\epsilon=0.5$,
$f_{\nu\bar{\nu}}\sim0.005-0.01$ for $\dot{M}_{-1}\sim1$, $f_{k}\sim
f_{\rm rot}$, and
$\tau_{J}\sim7.6B_{0,16}^{-2}(P_{r}/1\textrm{ms})^{2}$
s,\footnote{The annihilation efficiency $f_{\nu\bar{\nu}}$ for a
black hole hyperaccreting disk is $\sim10^{-4}$ for
$\dot{M}=0.1M_{\odot}$ s$^{-1}$, and this efficiency can be higher
and up to $\sim10^{-3}$ for the neutron star disk without fields.
The value of $f_{\nu\bar{\nu}}$ can be even higher for a magnetized
disk and funnel accretion as mentioned above. We take
$f_{\nu\bar{\nu}}\sim0.005-0.01$. However, the $f_{\nu\bar{\nu}}$
value is insensitive to the final results, since the period
$P_{r}\propto f_{\nu\bar{\nu}}^{-1/4}$. The spin-down timescale can
be approximately taken as the one in Thompson et al. (2004) and
Metzger et al. (2007), i.e., $\tau_{J}\propto(P_{r}/B_{0})^{2}$ for
$r_{A}<c/\Omega$.} then the critical value of the period $P$ for the
ratio $\dot{E}_{\rm anni}/\dot{E}_{\rm mag}$ being less than unity
is roughly $P_{r}\leq4$ ms. The critical value $\sim4$ ms will even
decrease if the disk accretion rate is higher with a higher
annihilation efficiency $f_{\nu\bar{\nu}}$, or the spin-down
timescale is longer. Keep in mind that this critical value is less
than $P_{rc}$ in equation (\ref{grb1}), which shows the wind to be
magnetically-dominated and relativistic without neutrino
annihilation. Therefore, in the extreme case of a millisecond
magnetar, the magnetically-dominated jet extracts the rotation
energy as the main energy source. If the stellar spin period is
around tens to hundreds of milliseconds, the polar jet can be feeded
by neutrino annihilation and magnetic energy together, and has a
bulk Lorentz factor even higher than $\sigma$. On the other hand,
for the magnetar period $P_{r}\geq P_{rc}$ in equation
(\ref{grb1}),\footnote{The maximum initial spin period of newborn
pulsars is still under debate. For example, Spruit \& Phinney (1998)
considered that the newborn pulsar without ``kicks" can reach a long
initial period $\sim100$ s. However, Heger et al. (2005) showed that
the spin period of the pulsar formed a the $10-15M_{\odot}$ star is
$10-15$ ms based on their stellar evolution code.} the jet formation
process goes back to be thermally-driven and feeded by the
annihilation process as discussed in paper II.\footnote{We mention
that Harikae et al. (2009) have discussed the combination effects of
MHD mechanism and annihilation process in the accreting black hole
system based on the collapsar scenario. They showed that the
collapsar with an initial field $\sim10^{10}$ G and angular momentum
of 1.5 times the angular momentum of the last stable orbit can
finally produce the MHD outflows and obtain the strong neutrino
heating in the polar funnel at the same time. However, how the
collapsar can generate a initial field with $\sim10^{10}$ G remains
an open question.}

\subsection{Disk Nucleosynthesis and $r$-Process Nucleosynthesis}

Besides the GRB or GRB-like (e.g., X-ray flashes) phenomena with
their associated supernovae, the hyperaccreting mangetar system can
also generate enough $^{56}$Ni and other elements from the disk for
the supernova (MacFadyen \& Woosley 1999; Kohri et al. 2005;
Nagataki 2006, 2007), and produce more heavier elements in the
neutrino-driven wind via $r$-process nucleosynthesis.

Simulation works based on the collapsar scenario (MacFayden \&
Woosley 1999; Nagataki et al. 2006, 2007) showed that $^{56}$Ni
produced in a jet of the collapsar is not sufficient to explain the
observed amount in a supernova with a duration about $\sim10$ s
(e.g., Mazzali et al. 2006; Soderberg et al. 2008). As a result, the
majority of $^{56}$Ni with mass $\sim10^{-2}M_{\odot}$ might be
synthesized in the accretion disk or in the disk outflows. Nagataki
et al. (2007) calculated that the $^{56}$Ni synthesized in the disk
can reach $\sim10^{-3}M_{\odot}$, but it should be carried out by a
later phase outflow or the viscous induced outflow. Also, Kohri et
al. (2005) discussed that the recombination process of nucleons
($n,p$) into nuclei can happen in the outflow and release an energy
about 8 MeV per nucleon. In the hyperaccreting magnetar scenario, on
the other hand, since the disk becomes hotter, denser with higher
pressure with a stronger field, the ejected $^{56}$Ni via the disk
outflow should be more than that estimated in the some region of the
collapsar without fields. However, if the magnetar field strength is
high enough and the funnel accretion is obvious, the disk region
outside the Alfv\'{e}n radius $r_{A}$ will be relatively small and
ejected even less $^{56}$Ni. Also, in the co-rotation region without
viscous heating at radius less than $r_{A}$, the disk will be cooled
down and produce much less $^{56}$Ni.

The main properties to determine the $r$-process production are the
asymptotic wind entropy $S^{a}$, asymptotic electron fraction
$Y_{e}^{a}$ and dynamical timescale $t_{\rm dyn}$. In general,
higher wind entropy, lower electron fraction and short dynamical
timescale are more favorable to produce heavier nuclei with higher
maximum $A$ number (Meyer \& Brown 1997; see also Thompson 2003).
Although the winds are thought to be a candidate for the
$r$-process, recent models without fields showed that the winds are
difficult to produce robust $r$-process nucleosynthesis for the
``classical" neutron star with a 1.4$M_{\odot}$ mass and 10 km
radius (e.g., Qian \& Woosley 1996; Thompson et al. 2001). Thompson
(2003) roughly estimated the $r$-process in the strong field
environment with $B_{0}\geq6\times10^{14}$ G. Since the strong field
can trap the wind in the neutrino heating region, the amplification
of the trapping timescale can lead to the amplification of the
entropy. This amplification may be sufficient to yield robust
third-peak $r$-process nucleosynthesis. In addition, Metzger et al.
(2007) showed that the presence of a magnetar field
$\sim2\times10^{14}$ G and mildly rapid rotation $\sim10$ ms moves
the ratio $(S^{a})^{3}/t_{\rm dyn}$, which is the critical wind
parameter to determine the condition for an $r$-process, an order of
magnitude more favorable for third-peak $r$-process nucleosynthesis.
Therefore, the magnetic fields from the central magnetar may play
significant role in inducing a successful strong $r$-process
nucleosynthesis. However, the problem is that, all of the works
considering magnetic fields do not include the strong field quantum
effects on the state of equations of the wind. In other words, they
do not consider the Landau level effect and the neutrino absorption
reaction rate modified by the strong fields. As discussed in Section
6.2, the strong field can decrease the neutrino absorption reaction
rate, and also decrease the mass loss rate of the neutrino-driven
wind as well. Also, the strong field can affect the density and
pressure distribution in the wind. Therefore, it is interesting to
consider whether the microphysical change in the strong field can
affect the final results of the $r$-process nucleosynthesis.
Furthermore, if we look into the scenario of hyperaccreting disks,
we should keep in mind that the disk will increase the wind entropy
and affect its distribution in the polar region where propagates the
stellar neutrino-driven wind (e.g., Nagataki et al. 2006, 2007).
Another possibility is that the disk outflows and the X-type winds,
rather than the stellar neutrino-driven winds, will also produce the
$r$-process elements with $A>130$, because the neutrino-to-proton
ratio in the disk and the outflows are sufficiently high, and the
mass loss rate from the disk is much higher than that from the
stellar surface (Kohri et al. 2005). However, whether the disk
outflows can be the site for $r$-process nucleosynthesis is still
under debate (Metzger et al. 2008a). Since the pressure of the
outflow will drop quickly above the disk and be less than the
magnetic pressure, the magnetic field above the disk may also play
significant role in increasing the outflow entropy as discussed in
Thompson (2003). As a result, more work should be done to understand
the influence of a strong field on the $r$-process in the
hyperaccreting magnetar systems.

\subsection{A Unified Scenario of Collapse-related-GRB Models}

Let us compare the hyperaccreting neutron star/mangetar model with
the isolated magnetar model, both of which are proposed to be able
to produce GRB and GRB-like events. The hyperaccreting neutron
star/magnetar disks can form in collapsars or compact binary
mergers. In the collapsar scenario, rotation and magnetic fields
make the core be possible to collapse into a massive neutron
star/magnetar rather than a black hole. On the other hand, isolate
magnetars are proposed to form via the rotating Type-Ib/c
supernovae, the mergers of compact binaries, or the
accretion-induced collapse of white dwarfs. Therefore, the
progenitors of the hyperaccreting compact star model and the isolate
magnetar model are actually very similar with each other. The main
difference is the environment: the disk around the neutron
star/magnetar will produce significant phenomena which will not be
produced by the isolate magnetar. Let us see the case of massive
star collapse for example. The isolate magnetar forms after a
successful supernova explosion, and the supernova shock has created
a cavity around the magnetar before it drives magnetically-dominated
and Poynting-dominated winds. On the contrary, the unsuccessful or
weak ongoing shock leads to the hyperaccreting system. The disk
material around the neutron star/magnetar comes from the continuous
infalling stellar envelope, or the fallback of stellar material
which has been ejected by the shock but cannot reach sufficiently
high velocity to escape the gravitational potential of the core.
Kumar et al. (2008) discussed the accretion rate of different
accretion stages, in which the accretion rate is different depending
on which stellar zone is accreted.

We try to use a unified point of view to consider the outcomes of
the massive star collapse. If the core collapse can initiate a
successful supernova, then an isolated black hole or a rotating
magnetar can form in the supernova remnant. Vietri \& Stella (1998,
1999) discussed the possibility of a neutron star further losing its
angular momentum by magnetic dipole or gravitational waves radiation
during a long time and collapsing to an accreting black hole. In
addition, the magnetar model shows the possibility that a variety of
magnetic activities from the isolate magnetar can produce a GRB
explosion after the supernova for a short time. On the other hand,
if the outgoing shock formed during a core collapse cannot compete
the continuous accretion from the stellar envelope, the collapse
leads to the formation of a collapsar system, in which the type of
the central compact object depends on many factors such as rotation,
equations of state and so on. The black hole disk may produce a GRB
explosion via neutrino annihilation or MHD processes, but the
annihilation mechanism may not produce sufficient energy for
energetic GRBs. The neutron star disk will increase the annihilation
efficiency compared to its black hole counterpart, and the increased
efficiency will be even higher if the central star is a magnetar. In
the hyperaccreting magnetar system, the increased annihilation
process and the magnetically-driven pulsar wind can work together to
generate a more powerful jet than that generated by a single
mechanism. The outflows and magnetically-dominated winds from the
disk is possible to feed a late-disk-induced supernova associated
with the GRB. In a word, the unified scenario shows that the
outcomes have a closed relation to the initial stellar properties
and the core collapse process itself.

However, such a unified scenario as the GRB central engine candidate
cannot be confirmed directly. What we can observe is the GRB-related
properties such as photon lightcurve, spectrum, neutrino emission,
different bands of afterglows, which can be traced back to the
central engine and show evidence for the existence of a central
neutron star. We do not want to show further evidences here (see Dai
2004, Dai et al. 2006; Fan \& Xu 2006; Yu \& Dai 2007 for more
details). Finally we want to mention two points. One is that, the
intermediate case, i.e., the fallback material forms a normal disk
around a magnetar, has been discovered (Wang et al. 2006). Although
the disk radiated IR emission is very different to the
hyperaccreting disk, this discovery shows a link between isolate
magnetar and hyperaccreting magnetar in the core collapses. Another
point is that, since some stars that are less massive than
Wolf-Rayet stars are more likely to form neutron stars and
magnetars, we cannot preclude the possibility that the collapse of
stars with an intermediate mass can also lead to the hyperaccreting
neutron star/magnetar systems and produce the GRB-like phenomena. In
fact, some evidences show that the GRB may be associated with Type
II supernova (Germany et al., 2000; Rigon et al., 2003), although
the evidences are still controversial today.

\section{Conclusions}

The hyperaccreting neutron star or magnetar disk systems cooled via
neutrino emission can form by the mergers of compact star binaries
or the collapses of rotational massive stars. Strong fields of the
magnetar can play a significant role in affecting the disk
properties and even changing the accretion process. Our motivation
in this paper is to investigate the influence of such strong
magnetic fields on the disks, and discuss implications of the
magnetar disk systems for the GRB and GRB-like events. Our
conclusions are as follows.

(i) We consider the magnetar field has a dipolar vertical component
$B_{z}$. The differential rotation between the disk and the magnetar
will generate a toroidal field component $B_{\phi}$, as well as a
relatively weak radial component $B_{r}$. The generated field can
have an open or closed configuration, depending on the disk's
viscous turbulence, magnetic diffusivity, disk angular velocity as
well as twist limitation. Similar to pervious works, we use the
parameter $\beta$ to measure the strength of the toroidal field
(equations (\ref{fieldstru01}) and (\ref{fieldstru02})). The
generated large-scale disk field coupled with the accretion flows
will transfer the angular momentum in radial direction and heat the
disk via Joule dissipation together with viscous stress outside the
Alfv\'{e}n radius $r_{A}$. On the other hand, since the distribution
and energy of electrons change significantly in the strong field
environment, the disk pressure and a variety of neutrino cooling
processes will be different compared to the case without fields.
Therefore, we discuss the quantum effects of the strong fields on
the disk thermodynamic and microphysical processes in Section 2, and
list the MHD conservation equations to describe the behavior of the
large-scale magnetic field coupling in the disk in Section 3.

(ii) The quantum effects and field coupling in MHD equations play
two competitive roles in changing the disk properties, the former to
decrease the pressure, density and neutrino luminosity with
increasing field strength (Figure 4), while the latter to increase
them (Figure 5). However, in most cases the large-scale field
coupling is more significant than the microphysical quantum effect
(Figure 7). Moreover, strong fields will change the electron
fraction distribution $Y_{e}$ in the disk significantly. Larger peak
of electron chemical potential $\eta_{e}$ with more degeneracy
electron states is obtained by stronger fields, while the change of
$\eta_{e}$ becomes more obvious for stronger fields.

(iii) Similar to the neutron star disk without fields, the values of
density, pressure, temperature, neutrino luminosity and electron
chemical potential become higher for higher accretion rate, but the
electron fraction $Y_{e}$ decreases with increasing the accretion
rate.

(iv) The magnetized disk maintaining a plane geometry would be more
favorable for an open field configuration rather than a closed one.
However, we still consider the two cases for completeness (Figures 6
and 9). For the disk with an open field and $\dot{M}=0.1M_{\odot}$
s$^{-1}$, higher ratio of $\beta/s$ leads to higher density and
pressure in the entire disk plane, higher temperature and neutrino
luminosity in the inner part of the disk, as well as lower $Y_{e}$
in the outer part. Here $s$ is the ratio of disk angular velocity
and the Keplerian velocity. Also, electrons becomes more degeneracy
at $\sim20-40$ km for higher $\beta/s$. On the other hand, the disk
properties with a closed field not only depends on the values of
$\beta$ and $s$, but also the spin period of the central magnetar.
Shorter period of the central star decreases the disk density,
pressure, $\eta_{e}$, but increases $Y_{e}$ at the outer part of the
disk. The effects of spin period are only obvious for sufficient
field strength (e.g., $\sim10^{16}$ G for the accretion rate
$0.1M_{\odot}$ s$^{-1}$) from the central star.

(v) The accretion flow in the disk plane outside the Alfv\'{e}n
radius is viscously stable. However, whether the disk is thermally
stable depends on many factors such as the disk region, magnetic
field strength, disk angular velocity and accretion rate. Generally
speaking, the disk will be definitely thermally unstable if its
non-field disk counterpart with the same accretion rate is unstable.
The disk region can also be thermally unstable even its non-field
counterpart is stable, if the region is near the Alfv\'{e}n radius
where magnetic field plays a more important role in transferring the
angular momentum and heating the disk than the viscous stress.

(vi) The thermally-driven outflows can also exist in the magnetar
disk beyond the Alfv\'{e}n radius for the case of
$\dot{M}<1.0M_{\odot}$ s$^{-1}$. We assume the accretion rate as a
power law in radius for simplicity (Equation (\ref{wind01})). The
outflow will take away the disk angular momentum (Equation
(\ref{wind02})) and may provide energy for a supernova associated
with the GRB explosion. The disk density, pressure and neutrino
luminosity decrease with increasing the outflow strength. Also, the
ratio of magnetic and matter pressure $P_{\rm B}/P_{\rm matter}$ and
the thickness of the disk become larger for stronger outflows.
However, since the total energy taken by the thermal outflow depends
on the size of the disk plane outside the Alfv\'{e}n radius, the
total energy injection rate from the outflow decreases significantly
for stronger stellar fields (Table \ref{tab01}). Besides the
thermally-driven outflow, strong fields inside the Alfv\'{e}n radius
lead the accretion flow to co-rotating with the stellar field, which
is possible to launch MHD winds along the field lines, and generate
the X-type wind near the disk-magnetosphere boundary. These
magnetically-dominated winds are nonrelativistic and may provide
energy for a supernova explosion.

(vii) In the hyperaccreting disks, the funnel accretion can only be
important for extremely strong fields, which depend on the accretion
rate. The accretion process along the disk plane will be truncated
in the stellar magnetosphere, so most of the accretion flow can be
approximately considered as being lifted along the closed field
lines onto the magnetar surface. In most cases, the hyperaccreting
infalling funnel flow is unlikely to develop a shock with Mach
number greater than unify (equation (\ref{f08})). The flow is
accelerated by the gravitational force and transfers the
gravitational binding energy to the kinematic energy and next to the
heating energy near the magnetar surface. The funnel flow will cover
a ring-like belt of ``hot spot" around the magnetar surface and emit
thermal neutrinos. Because the temperature is higher and neutrino
emission region is more concentrated than the hyperaccreting neutron
star without fields, the funnel accretion can accumulate more
powerful neutrino annihilation luminosity (Figure 13).

(viii) The neutrino annihilation process both from the magnetar
surface and from the disk plane will be higher than that without
fields. Moreover, the neutrino annihilation mechanism and the
magnetic activity from the stellar surface (i.e., the pulsar wind
mechanism) can work together to generate and feed an
ultra-relativistic jet along the stellar magnetic poles. If the
stellar spin period is sufficiently short (e.g., $\sim4$ ms for the
field $\sim10^{16}$ G and $\dot{M}=0.1M_{\odot}$ s$^{-1}$), the jet
from the magnetar will be magnetically-dominated and mainly feeded
by extraction the stellar rotational energy. If the magnetar spin
period is long (i.e., longer than the critical value $P_{rc}$ in
Equation (\ref{grb1})), the jet is thermally-driven and feeded by
the annihilation process. In the intermediate case, on the other
hand, the relativistic jet can be launched by the pulsar-wind-like
process and neutrino annihilation together.

\section*{Acknowledgements}
We would like to thank the anonymous referee for his/her very useful
comments that have allowed us to improve our paper. We also thank
Bing Zhang, Todd Thompson and Yi Xie for the helpful discussions and
for a critical reading of the text. D.Z. is supported by a
Distinguished University Fellowship from the Ohio State University.
Z.G.D. is supported by the National Natural Science Foundation of
China (grant 10873009) and the National Basic Research Program of
China (973 program) No. 2007CB815404.

\appendix

\section{Disk Equations without Field}

The thermodynamical, microphysical and conservation equations in a
hyperaccreting disk without fields can be seen in many previous
works. In order to compare them with the case with strong fields, we
list them systematically in this appendix. Here we do not consider
equations of the neutron star inner disk with a self-similar
structure as discussed in paper I and paper II.

The electron/positron density number reads
\begin{equation}
n_{e^{\pm}}=\frac{(m_{e}c)^{3}}{\pi^{2}\hbar^{3}}\int^{\infty}_{0}
\frac{x^{2}dx}{e^{(m_{e}c^{2}\sqrt{x^{2}+1}\mp\mu_{e})/k_{B}T}+1},\label{a01}
\end{equation}
and the electron/positron pressure is
\begin{equation}
P_{e^{\pm}}=\frac{1}{3}\frac{m_{e}^{4}c^{5}}{\pi^{2}\hbar^{3}}
\int^{\infty}_{0}\frac{x^{4}}{\sqrt{x^{2}+1}}\frac{dx}{e^{(m_{e}c^{2}\sqrt{x^{2}+1}\mp\mu_{e})/k_{B}T}+1}.\label{a02}
\end{equation}
The total pressure is the summation of the pressure of electrons,
nucleons, radiation and neutrinos (without magnetic pressure):
\begin{equation}
P=P_{e^{-}}+P_{e^{+}}+P_{\rm nuc}+P_{\rm rad}+P_{\nu},\label{a03}
\end{equation}
where $P_{\rm nuc}=\rho RT$ and $P_{\rm rad}=a_{B}T^{4}/3$ with $R$
being the gas constant and $a_{B}$ being the radiation constant.

The total neutrino cooling rate is the same as equation
(\ref{coolingrate1}) with various absorption and scattering depths
showed in the beginning of Section 2.2.

The neutrino cooling by electron-positron captures by nucleons are
contributed by three terms as follows:
\begin{equation}
\dot{q}_{p+e^{-}\rightarrow n+\nu_{e}}=\tilde{K}n_{p}\int_{q}^{\infty}\varepsilon(\varepsilon^{2}-1)^{1/2}(\varepsilon-q)^{3}f_{e^{-}}\textrm{d}\varepsilon,\label{a04}
\end{equation}
\begin{equation}
\dot{q}_{n+e^{+}\rightarrow p+\bar{\nu}_{e}}=\tilde{K}n_{n}\int_{1}^{\infty}\varepsilon(\varepsilon^{2}-1)^{1/2}(\varepsilon+q)^{3}f_{e^{+}}\textrm{d}\varepsilon,\label{a05}
\end{equation}
\begin{equation}
\dot{q}_{n\rightarrow p+e^{-}+\bar{\nu}_{e}}=\tilde{K}n_{n}\int_{1}^{q}\varepsilon(\varepsilon^{2}-1)^{1/2}(q-\varepsilon)^{3}(1-f_{e^{-}})\textrm{d}\varepsilon,\label{a06}
\end{equation}
when the field is absent or can be ignored. Here $\tilde{K}$,
$\varepsilon$, $q$ and $f_{e^{\mp}}$ as in Equations
(\ref{coolingrate3}) to (\ref{coolingrate5}) in Section 2.2. Other
cooling rates without fields have been showed in Section 2.2.

The chemical equilibrium is
\begin{equation}
\textrm{ln}\left(\frac{n_{n}}{n_{p}}\right)
=f(\tau_{\nu})\frac{2\mu_{e}-Q}{k_{B}T}+[1-f(\tau_{\nu})]\frac{\mu_{e}-Q}{k_{B}T},\label{a07}
\end{equation}
where the weight factor $f(\tau_{\nu})=\textrm{exp}(-\tau_{\nu})$
combines the formula from the neutrino-transparent limit case with
the neutrino-opaque limit case of the $\beta$-equilibrium
distribution, and $Q=q m_{e}c^{2}$.

A set of conservation equations without fields are as follows:

Mass conservation (continuity) equation reads
\begin{equation}
\dot{M}=-2\pi r\Sigma v_{r},\label{a08}
\end{equation}
this equation keeps the same with and without fields in the
vertically-integrated disks. However, we should consider another
continuity equation for outflows and winds as
\begin{equation}
\dot{M}=-4\pi r^{2}\rho v_{r}(\Delta\Omega/\Omega),\label{a09}
\end{equation}
where $\Delta\Omega$ is the opening solid angle of the wind.

The integrated angular momentum conservation equation is
\begin{equation}
\frac{\dot{M}}{3\pi}f=\nu\Sigma,\label{a10}
\end{equation}

The local energy conservation without fields is
\begin{equation}
Q_{\rm vis}^{+}=Q_{\rm adv}^{-}+Q_{\nu}^{-},\label{a11}
\end{equation}
where the viscosity heating term is
\begin{equation}
Q^{+}=\frac{3GM\dot{M}}{8\pi r^{3}}f\label{a12}
\end{equation} and the advection term is
\begin{equation}
Q_{\rm
adv}^{-}=v_{r}T\frac{\Sigma}{2r}\left[\frac{R}{2}\left(1+Y_{e}\right)+\frac{4}{3}g_{*}\frac{aT^{3}}{\rho}
\right],\label{a13}
\end{equation}
with $g_{*}=2$ for photons and $g_{*}=11/2$ for a plasma of photons
and relativistic $e^{-}e^{+}$ pairs.

The charge conservation equation is
\begin{equation}
\frac{\rho Y_{e}}{m_{B}}=n_{p}=n_{e^{-}}-n_{e^{+}},\label{a14}
\end{equation}
which will not be changed in strong fields if we do not consider
other charged particles created in the strong field environment.

\section{Solid Angle of Funnel Flow}
We take two methods to estimate the latitude and area of the
ring-like belt of the ``hot spot" on the magnetar surface. The first
one is just to follow the approximation made in Section 5.2. We
assume that the funnel accretion flow is lifted in the region with
radius between $r_{m}$ and $r_{A}$, and the disk plane will be
truncated at the magnetosphere $r_{m}$. Taking $r_{m}=r_{A}-\delta$,
we have the field line equation from the two disk radius  $r_{A}$
and $r_{m}$ as
\begin{equation}
r_{l1}=r_{A}\frac{\textrm{sin}^{2}\theta}{\textrm{sin}^{2}\Theta},\label{b01}
\end{equation}
\begin{equation}
r_{l2}=(r_{A}-\delta)\frac{\textrm{sin}^{2}\theta}{\textrm{sin}^{2}\Theta},\label{b02}
\end{equation}
or we obtain
\begin{equation}
\textrm{sin}\theta_{1}=\left(\frac{r_{*}}{r_{A}}\right)^{1/2}\textrm{sin}\Theta,\label{b03}
\end{equation}
\begin{equation}
\textrm{sin}\theta_{2}=\textrm{sin}(\theta_{1}+\Delta\theta)=\left(\frac{r_{*}}{r_{A}-\delta}\right)^{1/2}\textrm{sin}\Theta,\label{b04}
\end{equation}
If $\delta\ll r_{A}$, $\textrm{sin}\Theta\simeq1$, we can obtain an
analytic solution of $\Delta\theta$ and $\Delta\Omega/2\pi$. We have
\begin{equation}
\textrm{sin}\theta_{2}=\textrm{sin}\theta_{1}+\textrm{cos}\theta_{1}\Delta\theta=\left(\frac{r_{*}}{r_{A}}\right)^{1/2}\left(1+\frac{\delta}{2r_{A}}\right),\label{b05}
\end{equation}
or
\begin{equation}
\Delta\theta=\frac{\delta}{2}\frac{1}{\sqrt{r_{A}(r_{A}-r_{*})}},\label{b07}
\end{equation}
The solid angle spanned by the hot spot ring-like belt can be
calculated as
\begin{equation}
\Delta\Omega=\int_{0}^{2\pi}d\phi\int_{\theta_{1}}^{\theta_{2}}\textrm{sin}\theta d\theta\label{f05}
\end{equation}
and we obtain an analytic result of the ratio $\Delta\Omega/2\pi$ as
a function of $r_{A}$ and $\delta$.
\begin{equation}
\frac{\Delta\Omega}{2\pi}=\frac{2\pi \textrm{sin}\theta_{1}\Delta\theta}{2\pi}=\frac{\delta}{2r_{A}}\sqrt{\frac{r_{*}}{r_{A}-r_{*}}},\label{b08}
\end{equation}
We do not want to solve detailed conservation equations or a set of
MHD differential equations as in Ustyugova et al. (1999). Table
\ref{tab02} gives the results with different parameters of
$r_{A}/r_{*}$, $\delta/r_{*}$ and $\Theta$ as in the case of
hyperaccreting disks with strong fields. The typical value for
$\Delta\Omega/2\pi$ is around $10^{-2}$, except for the case when
$r_{A}$ is small and $\delta\sim r_{A}$, which can still be
considered as disk plane accretion. Stronger field makes the funnel
flow be more significant and the value $\Delta\Omega/2\pi$ be
smaller.

Next we give another scenario based on the consideration that the
disk has a thickness near the magnetosphere $r_{m}$. We try to give
a simple mathematical model: The ring-like belt is formed by the
accretion matter at $r_{m}$ with different height $z$ in the
vertical direction. In other words, the ring-like belt area on the
magnetar surface can be traced back to the disk plane vertical
section at $r_{m}$.
\begin{equation}
r_{l1}=r_{m}\frac{\textrm{sin}^{2}\theta}{\textrm{sin}^{2}\Theta},\label{b09}
\end{equation}
\begin{equation}
r_{l2}=r'\frac{\textrm{sin}^{2}\theta}{\textrm{sin}^{2}\Theta},\label{b10}
\end{equation}
where $r'$ satisfies
\begin{equation}
r'=\frac{r_{m}(1+H^{2}/r_{m}^{2})^{1/2}}{\textrm{cos}^{2}(H/r_{m})}\textrm{sin}^{2}\Theta,\label{b11}
\end{equation}
Similarly, if $\textrm{sin}\Theta\simeq1$ and $H^{2}/r_{m}^{2}\ll1$,
we have
\begin{equation}
\textrm{sin}\theta_{1}=\left(\frac{r_{*}}{r_{m}}\right)^{1/2}\textrm{sin}\Theta,\label{b12}
\end{equation}
\begin{equation}
\textrm{sin}\theta_{2}=\left(\frac{r_{*}}{r_{m}}\right)^{1/2}\frac{\textrm{cos}(H/r_{m})}{(1+H^{2}/r_{A}^{2})^{1/4}}\textrm{sin}\Theta,\label{b13}
\end{equation}
\begin{equation}
\Delta\theta=\frac{3}{4}\sqrt{\frac{r_{*}}{r_{m}-r_{*}}}\left(\frac{H}{r_{m}}\right)^{2},\label{b14}
\end{equation}
\begin{equation}
\frac{\Delta\Omega}{2\pi}=\frac{2\pi \textrm{sin}\theta_{1}\Delta\theta}{2\pi}=\frac{3}{4}\sqrt{\frac{r_{*}}{r_{m}}}
\sqrt{\frac{r_{*}}{r_{m}-r_{*}}}\left(\frac{H}{r_{m}}\right)^{2}
\end{equation}
We can take $r_{m}$ as a fraction of $r_{A}$. Table \ref{tab03}
gives numerical results in this case. The typical value of
$\Delta\Omega/2\pi$ is as $\sim10^{-3}-10^{-2}$, which can be
slightly smaller than that obtained using the first estimation.

On the other hand, the accretion rate onto the neutron star without
fields, as discussed in paper II, can be estimated as $(2\pi
r_{*}H)/(2\pi r_{*}^{2})=H/r_{*}$ (note that we consider the disk
over a half thickness). Thus the solid angle in this case is in the
order of $\geq0.1$, which is larger than the area formed by funnel
accretion.

\begin{table}
\begin{center} Disk Heating Rate and Maximum Outflow
Energy Rate
\footnotesize{
\begin{tabular} {rccc|ccc}
\hline\hline
&  & Disk Heating Energy & & & Max Outflow Energy & \\
&  & Rate ($10^{51}$ erg s$^{-1}$) & & & Rate ($10^{51}$ erg s$^{-1}$) & \\
\hline
& $B_{0}=10^{15}$ G & $B_{0}=10^{16}$ G & $B_{0}=10^{17}$ G & $B_{0}=10^{15}$ G & $B_{0}=10^{16}$ G & $B_{0}=10^{17}$ G\\
\hline
$\xi$=0.2 & 29.3 & 25.9 & 8.08 & 8.60 & 8.63 & 1.45\\
$\xi$=0.6 & 18.3 & 15.7 & 6.47 & 19.5 & 18.8 & 3.07\\
$\xi$=0.9 & 13.5 & 11.7 & 5.53 & 24.4 & 22.9 & 4.01\\
\hline \hline
\end{tabular}}
\caption{The disk heating rate is calculated in the region outside
the Alfv\'{e}n radius, where differential rotation and viscosity are
significant: $\dot{E}_{\rm heat}=\int_{r_{A}}^{r_{\rm out}}2\pi
rdr$. Here the outflow index $\xi=0.2, 0.6, 0.9$, and the accretion
rate $\dot{M}=0.5M_{\odot}$ s$^{-1}$. The max thermally-driven
energy rate is estimated using equation (\ref{wind03}). The total
thermal outflow energy can be considered as $0.1-1$ fraction of the
maximum energy, as discussed in paper II.}\label{tab01}
\end{center}
\end{table}

\begin{table}
\begin{center} Disk Properties Depending on Magnetar Field Strength \footnotesize{
\begin{tabular} {ll}
\hline\hline
$B_{0}\sim10^{14}$ G & no funnel accretion, disk is similar to that without fields\\
$B_{0}\sim10^{15}$ G & no funnel accretion, MHD coupling is important in the disk inner region, thermal outflow\\
$B_{0}\sim10^{16}$ G & weak funnel accretion, disk is denser, hotter
with higher pressure, brighter $L_{\nu\bar{\nu}}$ from the disk,\\
& thermal outflows from $r>r_{A}$, magnetic winds inside\\
$B_{0}\sim10^{17}$ G & strong funnel accretion, but no shock, much
brighter $L_{\nu\bar{\nu}}$ from the stellar ``hot spot",\\
& magnetically-dominated wind significantly\\
\hline \hline
\end{tabular}}
\caption{The accretion rate near the Alfv\'{e}n radius $r_{A}$ or
the stellar surface $r_{*}$ (depending on $r_{A}>r_{*}$ or
$r_{A}<r_{*}$ respectively) is $\dot{M}=0.1M_{\odot}$ s$^{-1}$. The
neutrino cooling emission is efficient in this case, and the
accretion flow is an NDAF. The inner region satisfying the
self-similar structure (papers I and II) can be ignored even the
disk is similar to that without fields.}\label{tab04}
\end{center}
\end{table}

\newpage
\begin{table}
\begin{center}
Values of $\Delta\Omega/2\pi$ on Magnetar Surface Based on the First
Scenario
\begin{tabular}{r|ccc|ccc}
\hline\hline
 &  & $r_{m}/r_{*}$($\Theta=90^{\circ}$) & & & $r_{m}/r_{*}$ ($\Theta=75^{\circ}$)& \\
$\delta/r_{*}$ & 2 & 4 & 7 & 2 & 4 & 7\\
\hline
0.1 & 1.84e-2 & 6.38e-3 & 2.74e-3 & 1.77e-2 & 6.16e-3 & 2.64e-3\\
0.5 & 0.110 & 3.46e-2 & 1.43e-2 & 0.106 & 3.34e-2 & 1.38e-2\\
1 & 0.292 & 7.74e-2 & 3.02e-2 & 0.282 & 7.48e-2 & 2.92e-2\\
2 & -- & 0.208 & 0.070 & -- & 0.200 & 6.70e-2\\
\hline \hline
\end{tabular}
\caption{The ratio of the solid angle of the ``hot spot" ring-like
belt $\Delta\Omega$ formed by funnel accretion to the half total
solid angle $2\pi$ based on the first scenario in Append B, in which
the accreted matter onto the ring-like belt of ``hot spot" is from
the region between the magnetosphere edge $r_{m}$ and the Alfv\'{e}n
radius $r_{A}$. We take $\delta/r_{*}$ from 0.1 to 2, and
$r_{m}/r_{*}$ from 2 to 7. Here the left three columns are for
$\Theta=90^{\circ}$ and the right three ones are for
$\Theta=75^{\circ}$. The typical value of $\Delta\Omega$ is around
$10^{-2}$.}\label{tab02}
\end{center}
\end{table}

\newpage
\begin{table}
\begin{center}
Values of $\Delta\Omega/2\pi$ on Magnetar Surface Based on the
Second Scenario
\begin{tabular}{r|ccc|ccc}
\hline\hline
 &  & $\Theta=90^{\circ}$ ($r_{A}/r_{*}$) & & & $\Theta=75^{\circ}$ ($r_{A}/r_{*}$)& \\
$H/r_{m}$ & 2 & 4 & 7 & 2 & 4 & 7\\
\hline
0.1 & 5.28e-3 & 3.74e-3 & 2.82e-3 & 5.50e-3 & 3.62e-3 & 2.72e-3\\
0.3 & 4.60e-2 & 3.26e-2 & 2.46e-2 & 4.44e-2 & 3.14e-2 & 2.38e-2\\
0.5 & 0.120 & 8.50e-2 & 6.42e-2 & 0.116 & 8.22e-2 & 6.20e-2\\
\hline \hline
\end{tabular}
\caption{The ratio of the solid angle of the ``hot spot" ring-like
belt $\Delta\Omega$ to the total solid angle $2\pi$ based on the
second scenario in Append B, in which the accreted matter onto the
ring-like belt of ``hot spot" is from the disk vertical section
surface at $r=r_{m}$. We take the disk thickness at this radius as
$H/r_{m}=0.1,0.3, 0.5$, and $r_{m}/r_{*}$ also from 2 to 7. The
typical value of $\Delta\Omega$ is still around $10^{-3}-10^{-2}$ in
this scenario.}\label{tab03}
\end{center}
\end{table}


\newpage
\begin{figure}
\resizebox{\hsize}{!} {\includegraphics{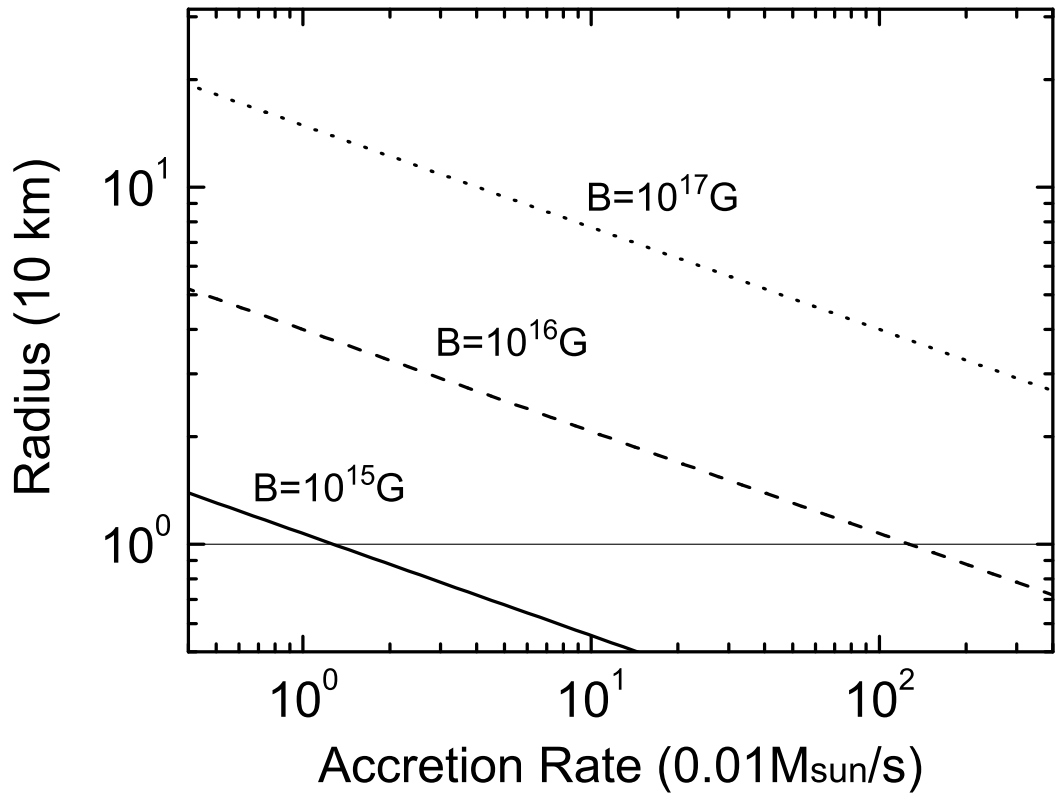}}
\caption{Alfv\'{e}n radius $r_{A}$ as a function of accretion rate
in hyperaccreting disks. We take the magnetar field as $10^{15},
10^{16}$  and $10^{17}$ G.}
\end{figure}
\newpage
\begin{figure}
\resizebox{\hsize}{!} {\includegraphics{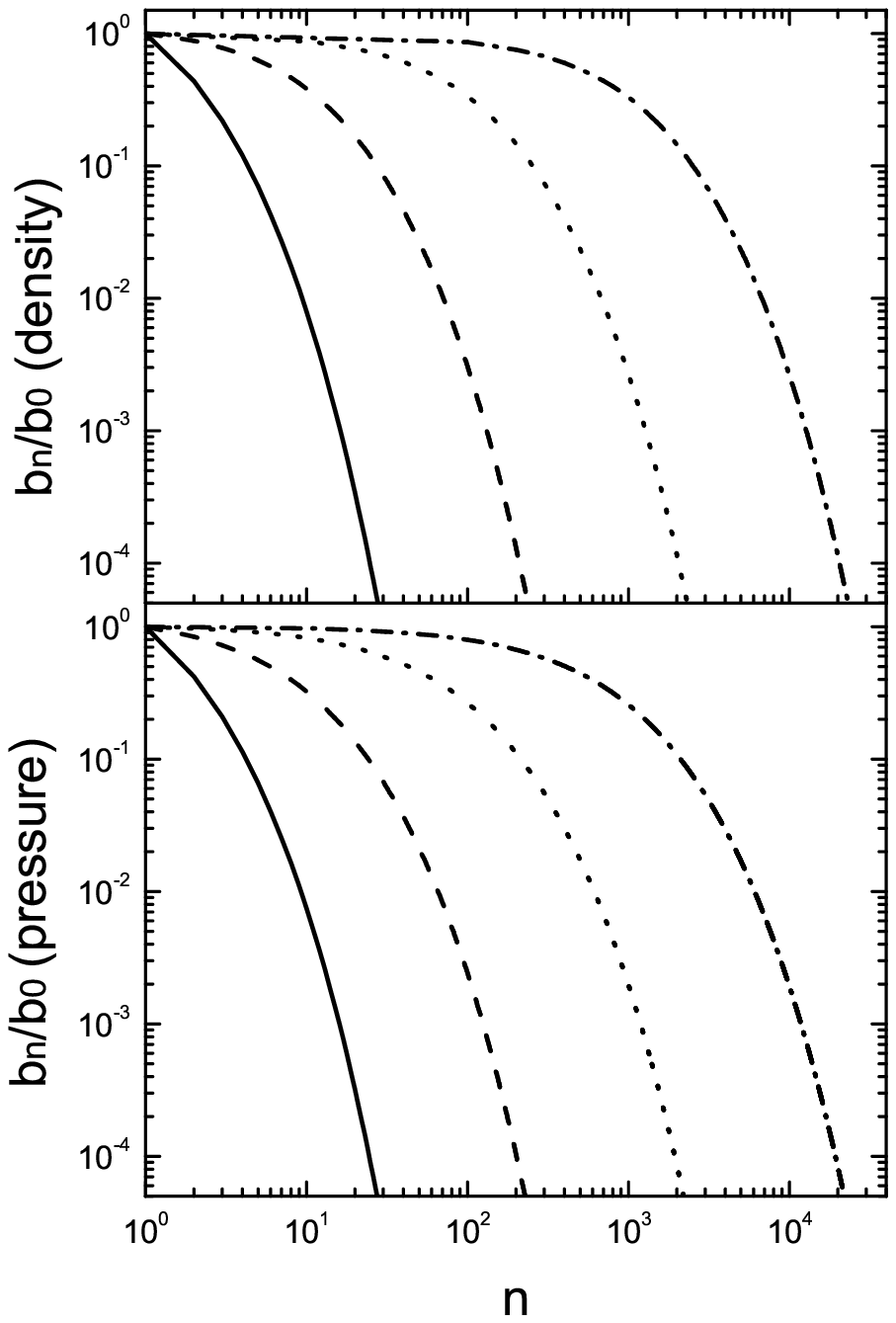}
\includegraphics{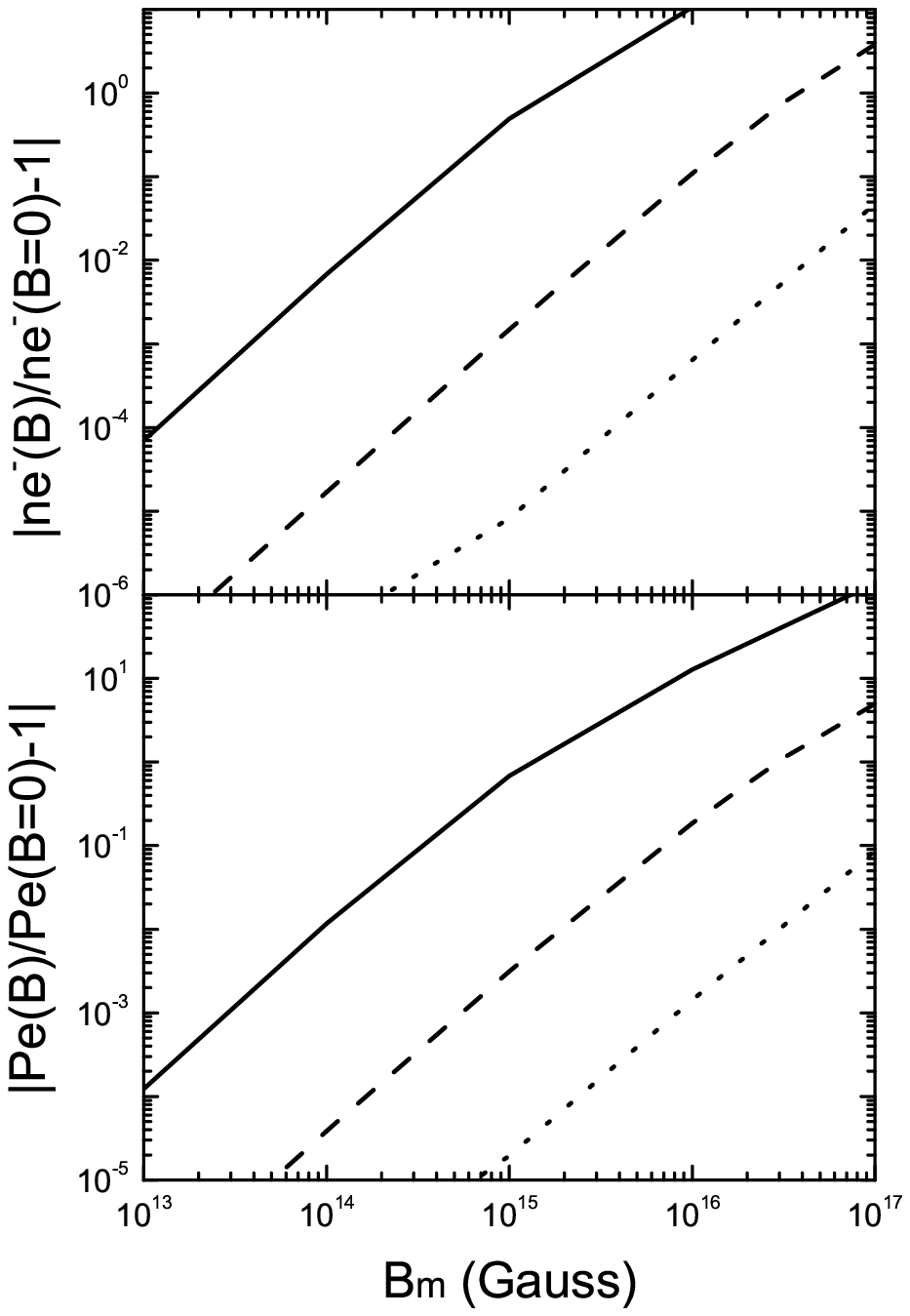}}
\caption{\textit{Left}: Convergence performance of Landau level
series of electron density and pressure with different magnetic
fields $B=10^{16}$ G ({\em solid line}), $10^{15}$ G ({\em dashed
line}), $10^{14}$ G ({\em dotted line}) and $10^{13}$ G ({\em
dash-dotted line}) and temperature $T=5\times10^{10}$ K and chemical potential $\eta_{e}=1$,
where $b_{n}$ is the $n$ term in the density and
pressure series (\ref{den01}) and (\ref{pre01}). \textit{Right}: Comparison of density and
pressure with and without magnetic fields with temperature $T=10^{10}$ K ({\em solid line}),
$5\times10^{10}$ K ({\em dashed line}) and $2\times10^{11}$ K ({\em dotted line}).}
\end{figure}

\newpage
\begin{figure}
\resizebox{\hsize}{!} {\includegraphics{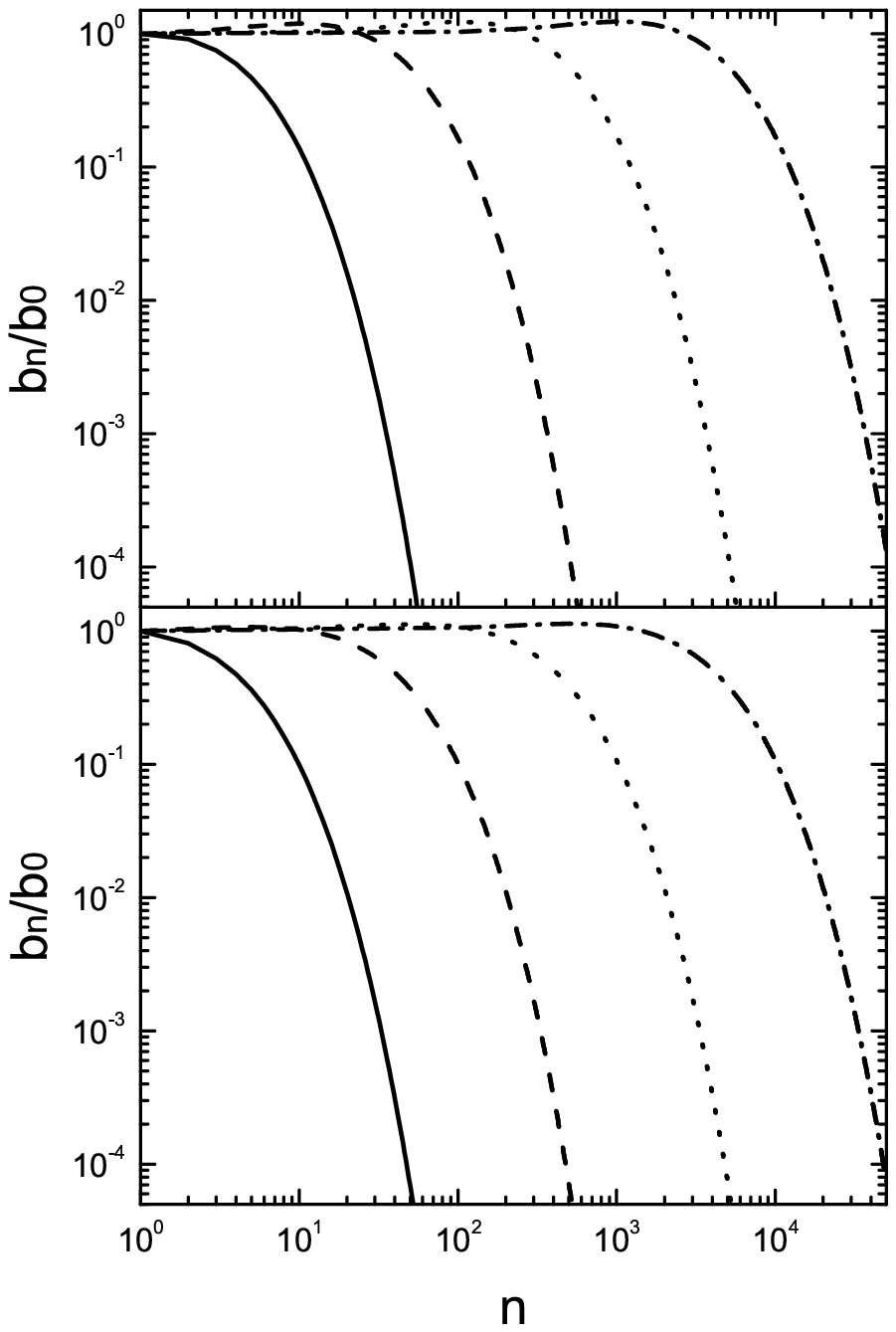}
\includegraphics{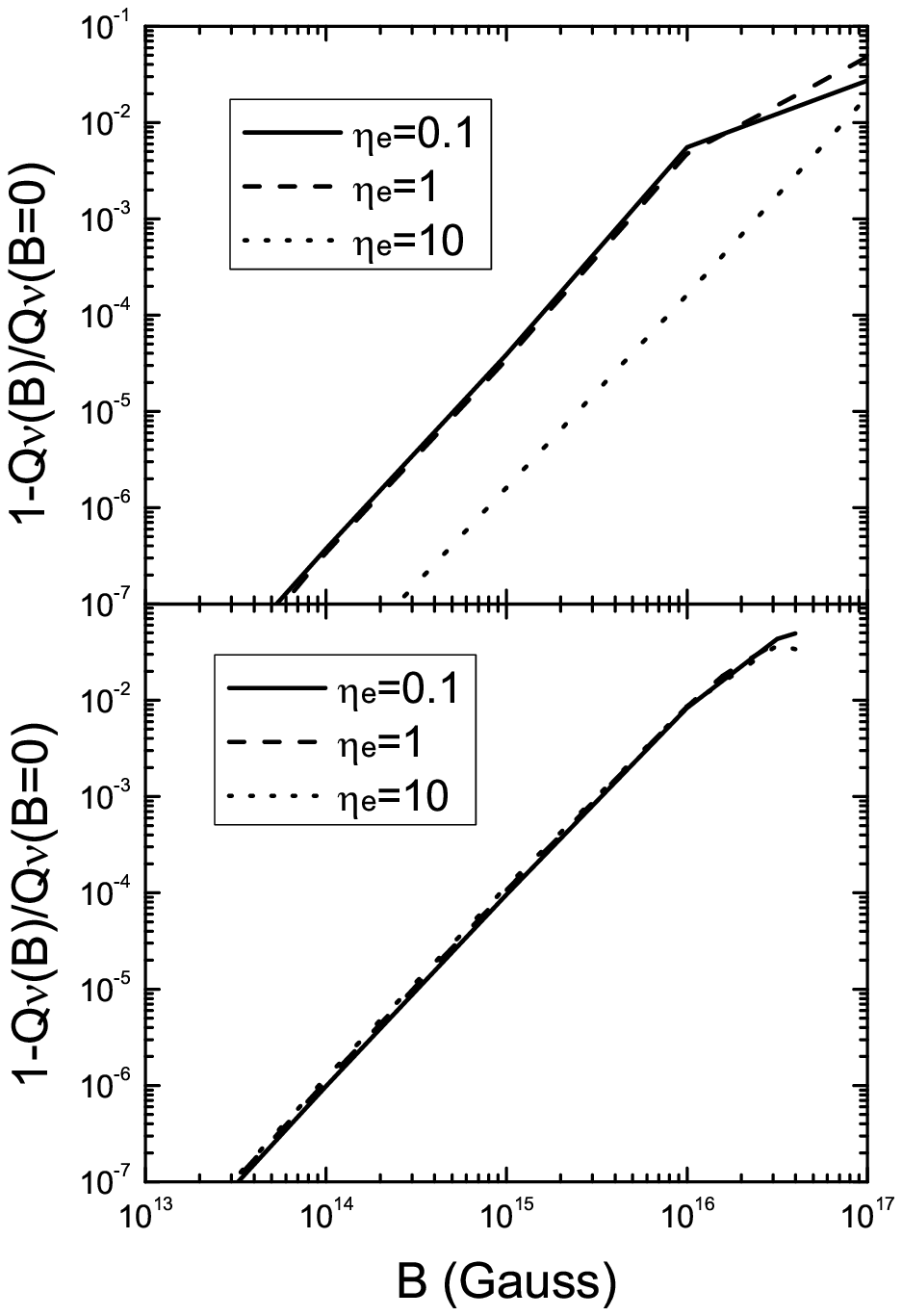}}\caption{\textit{Left}: Convergence performance
of Landau level series of neutrino cooling with different magnetic
fields. Upper one for the cooling rate $\dot{q}_{p+e^{-}\rightarrow
n+\nu_{e}}$ and lower for $\dot{q}_{n+e^{+}\rightarrow
p+\bar{\nu}_{e}}$, where temperature $T=5\times10^{10}$ K and
chemical potential $\eta_{e}=1$ is adopted. Lines are as in Fig. 1.
\textit{Right}: Comparison of neutrino cooling rates with and
without magnetic fields. Temperature $T=5\times10^{10}$ K and
$\eta_{e}=0.1,1$ and 10. Also, upper is for
$\dot{q}_{p+e^{-}\rightarrow n+\nu_{e}}$ and lower for
$\dot{q}_{n+e^{+}\rightarrow p+\bar{\nu}_{e}}$.}
\end{figure}

\newpage
\begin{figure}
\resizebox{\hsize}{!} {\includegraphics{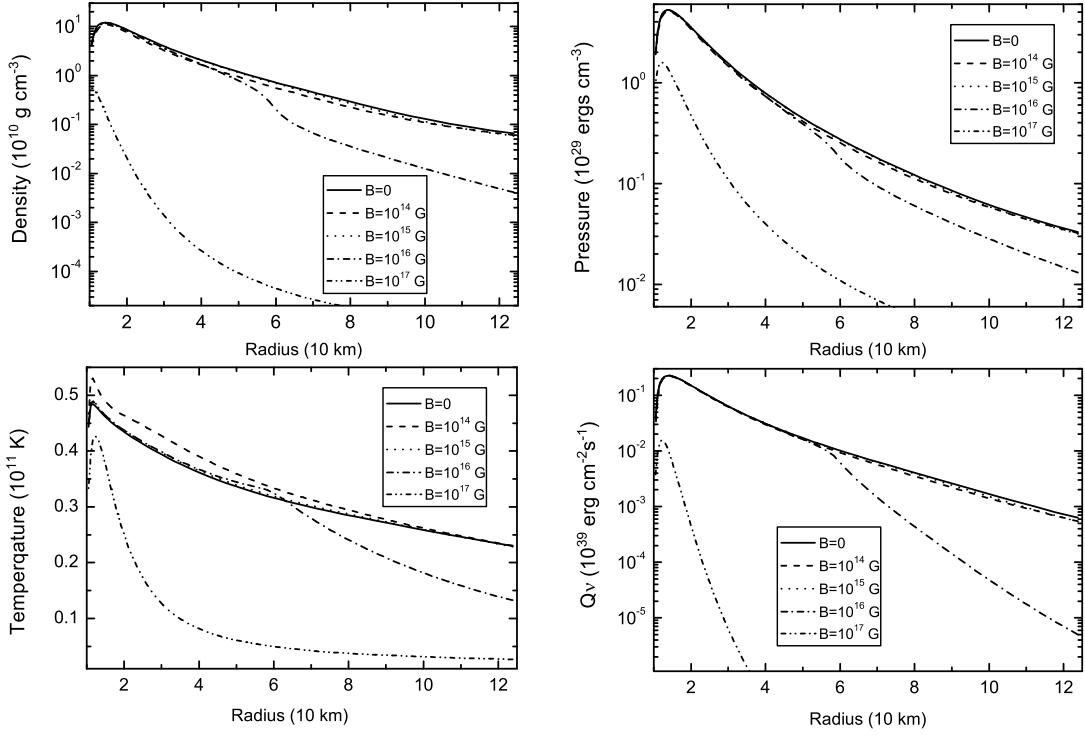}} \caption{Quantum
effects of microphysics and thermodynamics in strong magnetic fields
to affect the disk properties. In order to see the effects clearly,
we still take the non-magnetized conservation equations (i.e,
(\ref{a08}), (\ref{a10}) to (\ref{a14}) in Appendix A), and only
change a set of equations of state in Section 2. Also, we take the
magnetic field as uniform $B=10^{14}, 10^{15}, 10^{16}, 10^{17}$ G
and without field, the accretion rate $\dot{M}=0.1M_{\odot}$
s$^{-1}$.}
\end{figure}

\newpage
\begin{figure}
\resizebox{\hsize}{!} {\includegraphics{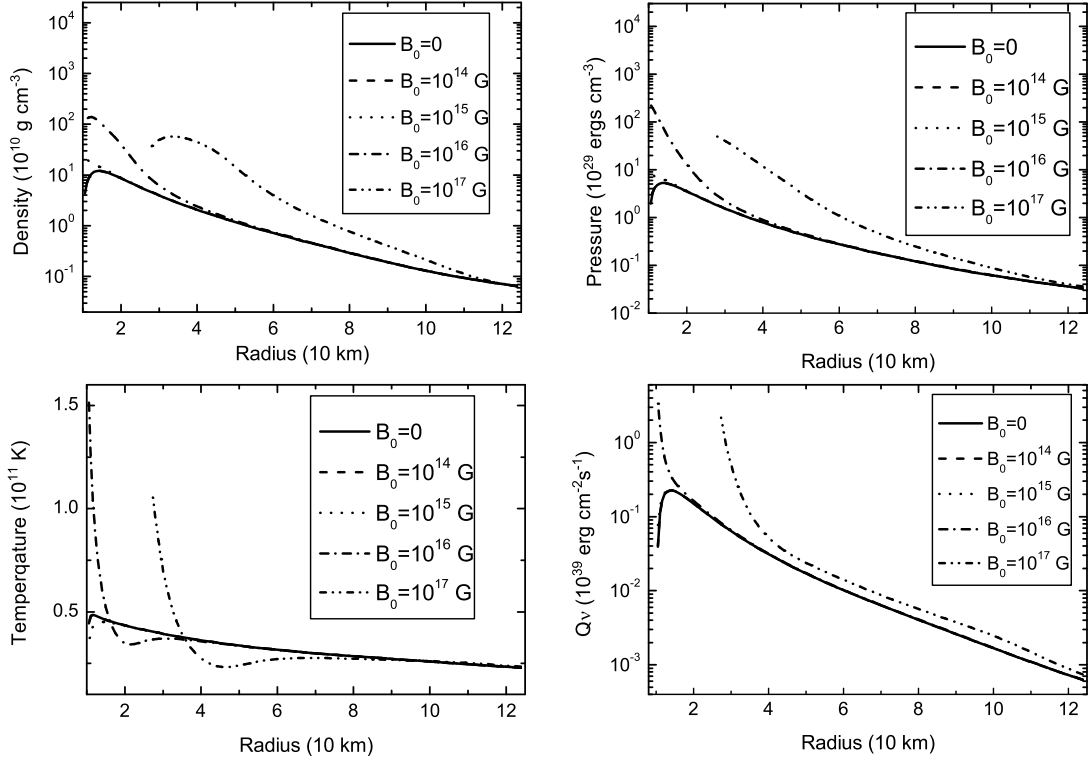}} \caption{Effects
of strong magnetic field coupling on angular momentum transfer and
heating via Joule dissipation on the disk. We do not consider the
field quantum effects as discussed in Section 2, but only adopt the
MHD equations in Section 3. The magnetic field topology is dipolar,
and an open configuration with parameter $\beta=0.6$, angular
velocity ratio $s=1$, and $B_{0}=10^{14}, 10^{15}, 10^{16}, 10^{17}$
G. The accretion rate is the same as in Figure 4.}
\end{figure}
\newpage
\begin{figure}
\resizebox{\hsize}{!} {\includegraphics{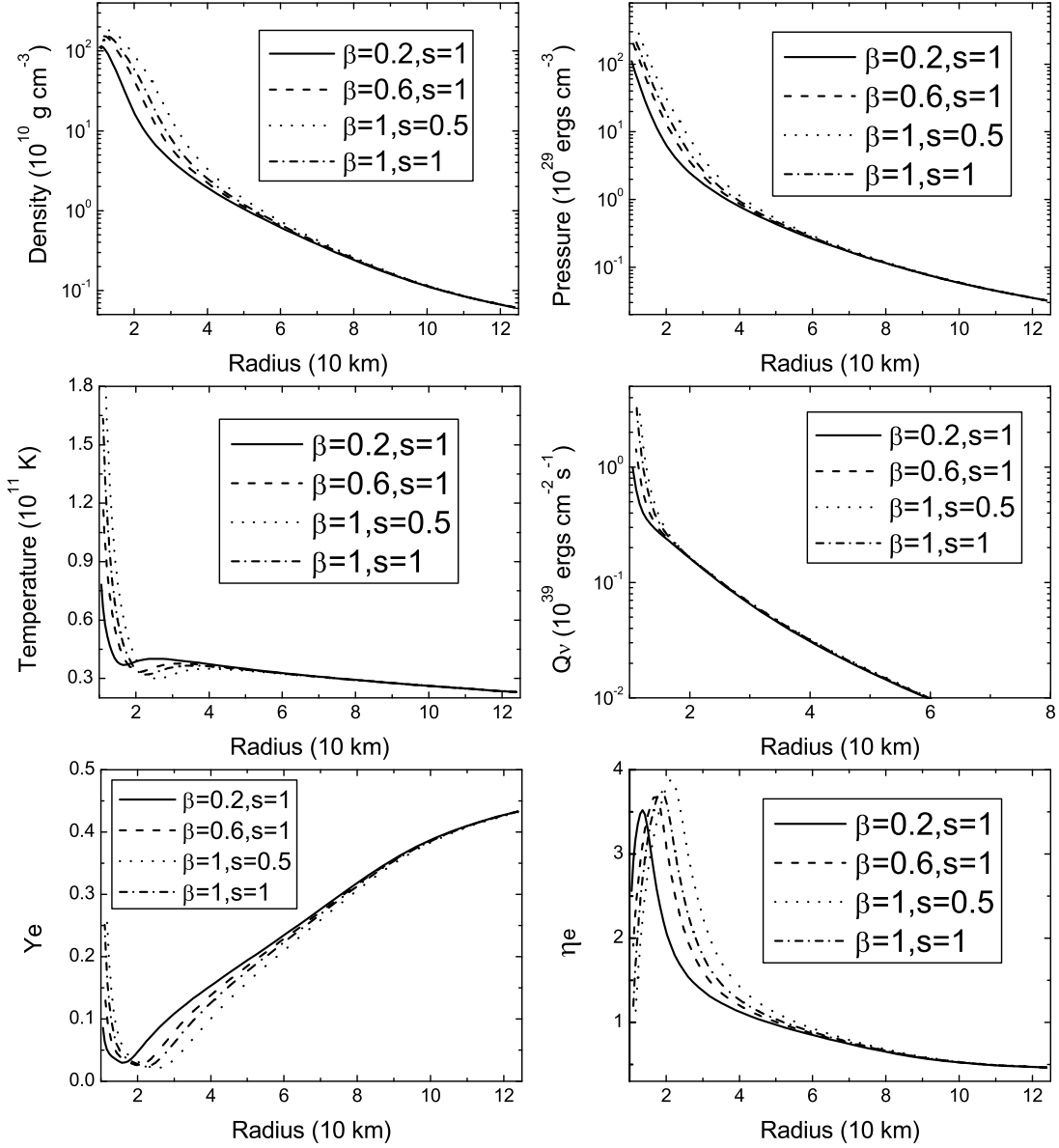}} \caption{Disk
structure with $\dot{M}=0.1M_{\odot}$ s$^{-1}$ for the magnetar
surface vertical field $B_{0}=10^{16}$ G. The magnetic field in the
disk is open with parameter $\beta=0.2,0.6, 1$ and disk angular
velocity $s=0.5, 1$.}
\end{figure}
\newpage
\begin{figure}
\resizebox{\hsize}{!} {\includegraphics{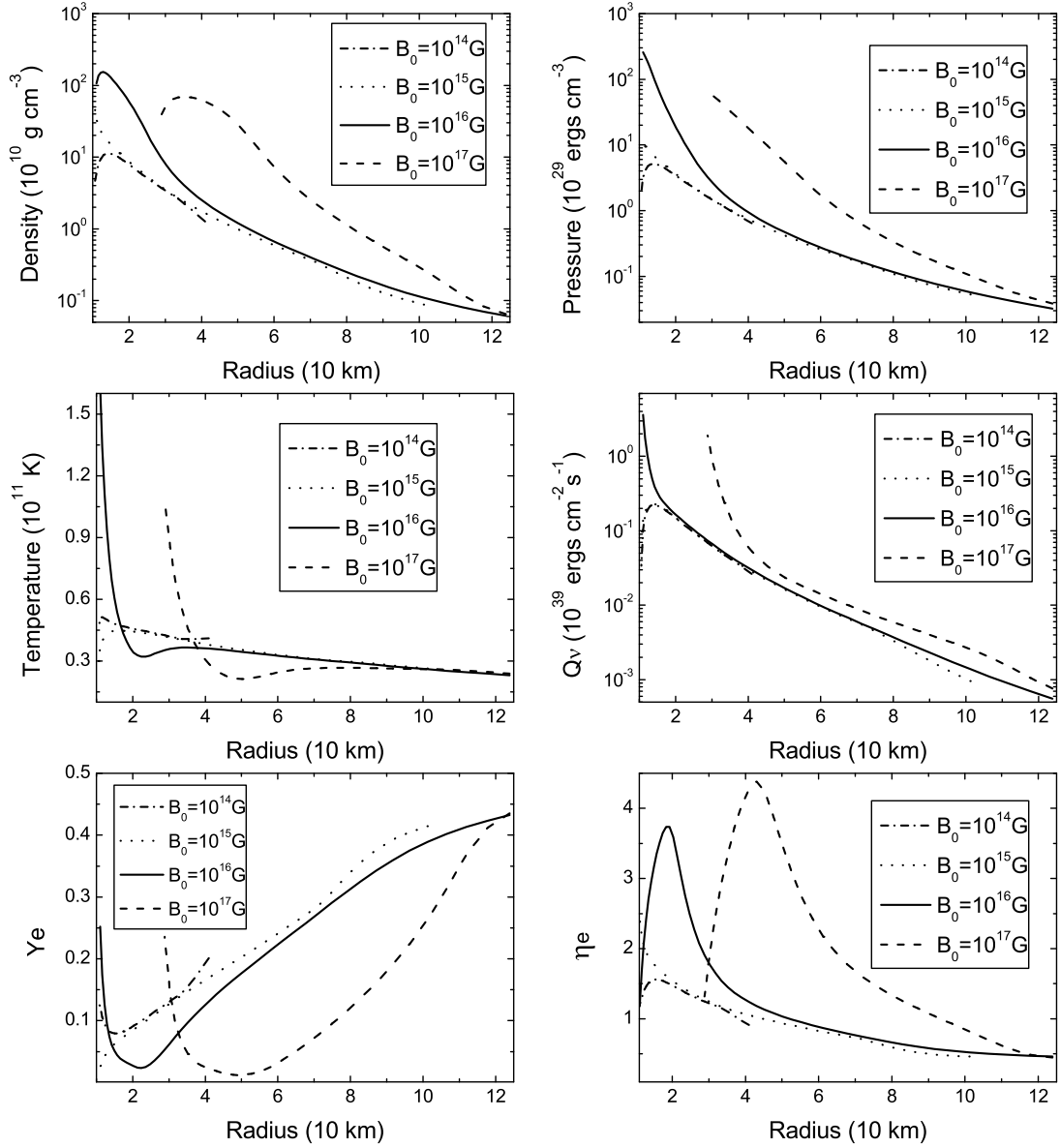}} \caption{Disk
structure with $\dot{M}=0.1M_{\odot}$ s$^{-1}$ for the magnetar
surface vertical field $B_{0}=10^{14},10^{15},10^{16}$ and $10^{17}$
G, where the magnetic field in the disk is in an open configuration
with $\beta=1$ and $s=1$.}
\end{figure}
\newpage
\begin{figure}
\resizebox{\hsize}{!} {\includegraphics{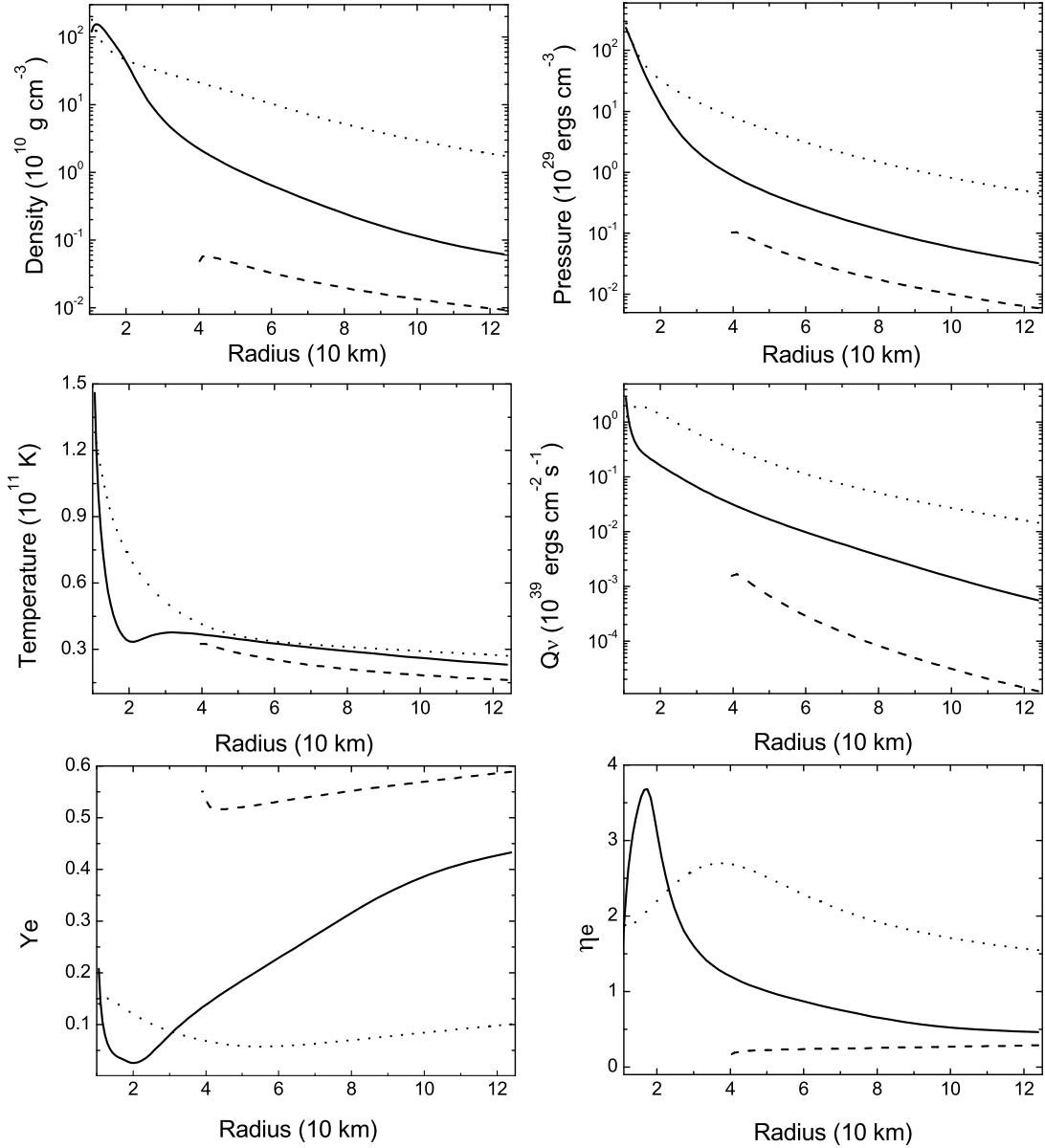}} \caption{Disk
structure for different disk accretion rate $\dot{M}=0.02$ ({\em
solid line}),0.1 ({\em dashed line}) and 1 $M_{\odot}$ s$^{-1}$
({\em dotted line}) with open disk field $\beta=0.6$ and $s=1$,
while the surface vertical field $B_{0}=10^{16}$ G.}
\end{figure}
\newpage
\begin{figure}
\resizebox{\hsize}{!} {\includegraphics{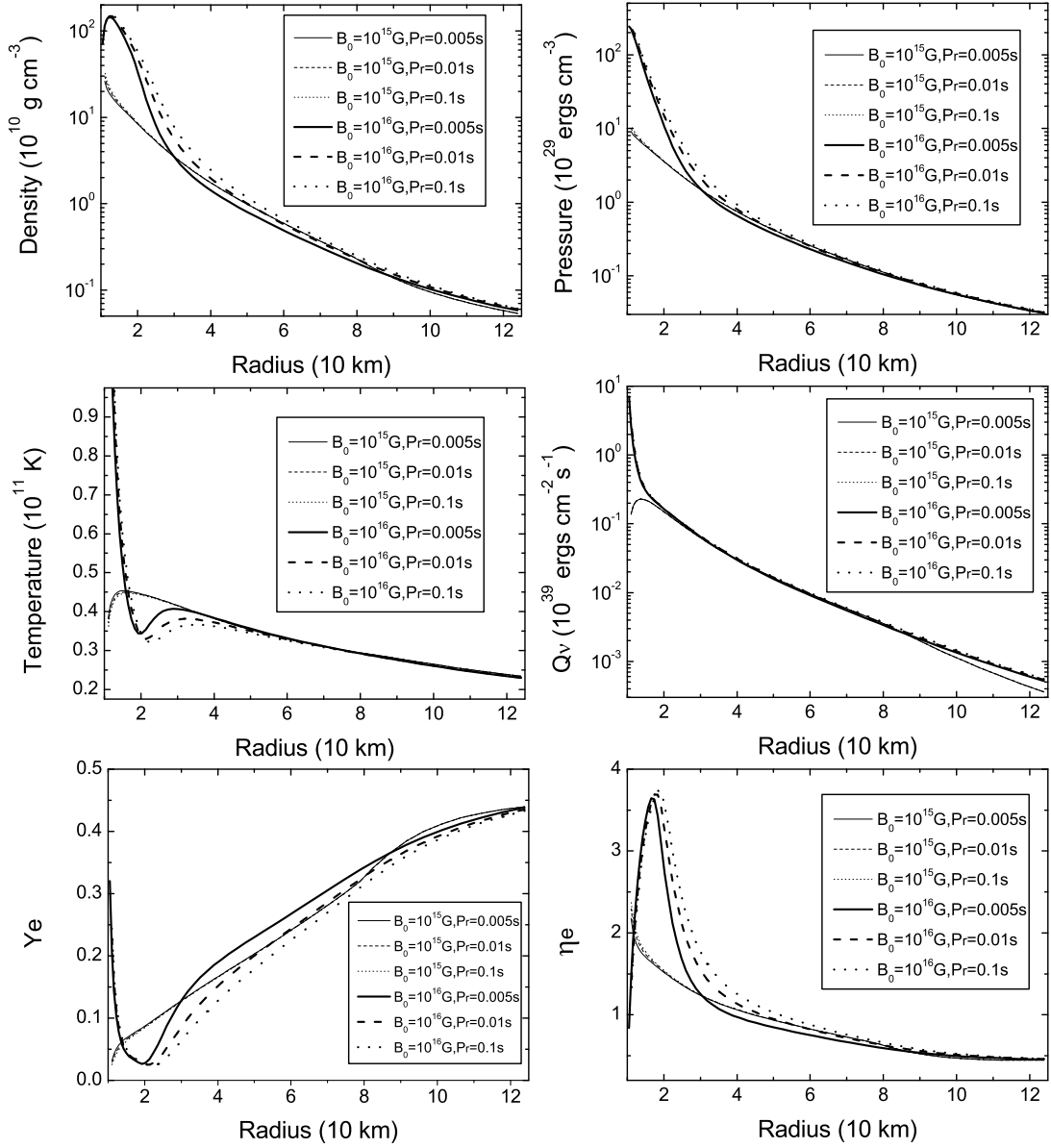}} \caption{Disk
structure with $\dot{M}=0.1M_{\odot}$ s$^{-1}$ for the surface
vertical field $B_{0}=10^{15}, 10^{16}$ G, the disk magnetic field
is closed with $\beta\simeq1$, $s=1$. The period of a central
magnetar $P_{r}=0.005,0.01$ and $0.1$ sec.}
\end{figure}
\newpage
\begin{figure}
\resizebox{\hsize}{!} {\includegraphics{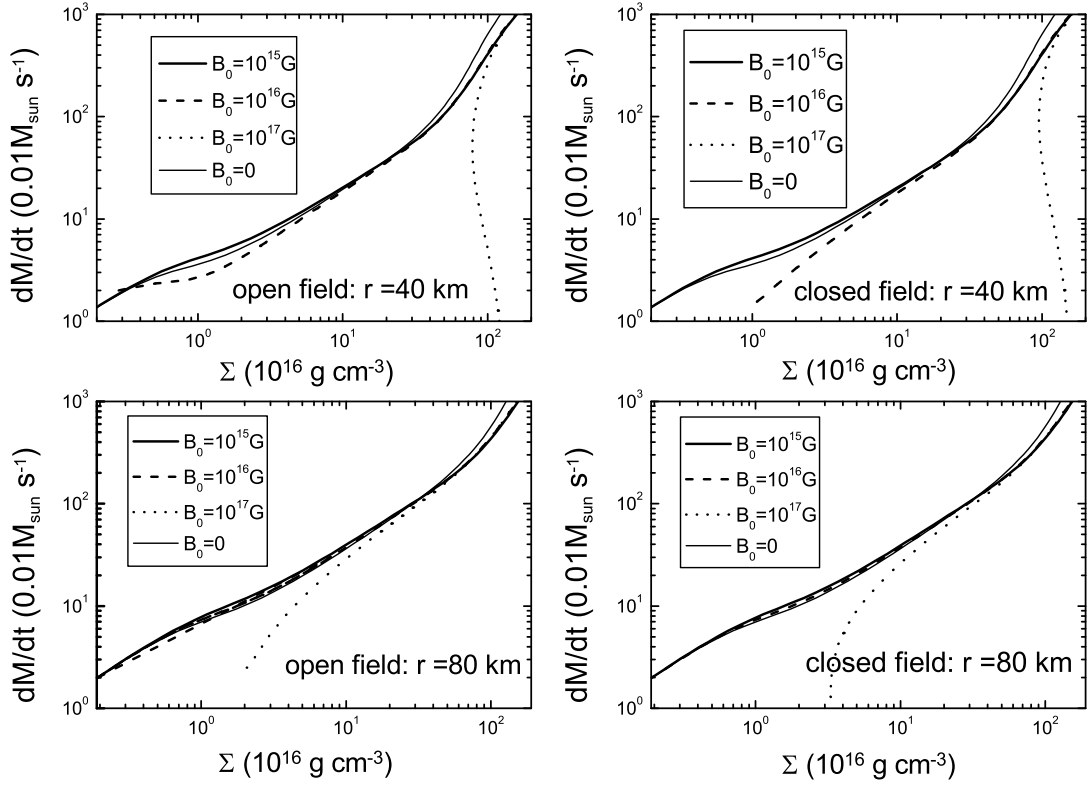}}
\caption{$\dot{M}-\Sigma$ curves for given disk radius $r=$40 and 80
km. In the case of an open disk field we take $\beta=0.6$ and $s=1$;
while in the case of a closed disk field we take $\beta=1$, $s=1$
and $P_{r}=100$ms.}
\end{figure}
\newpage
\begin{figure}
\resizebox{\hsize}{!} {\includegraphics{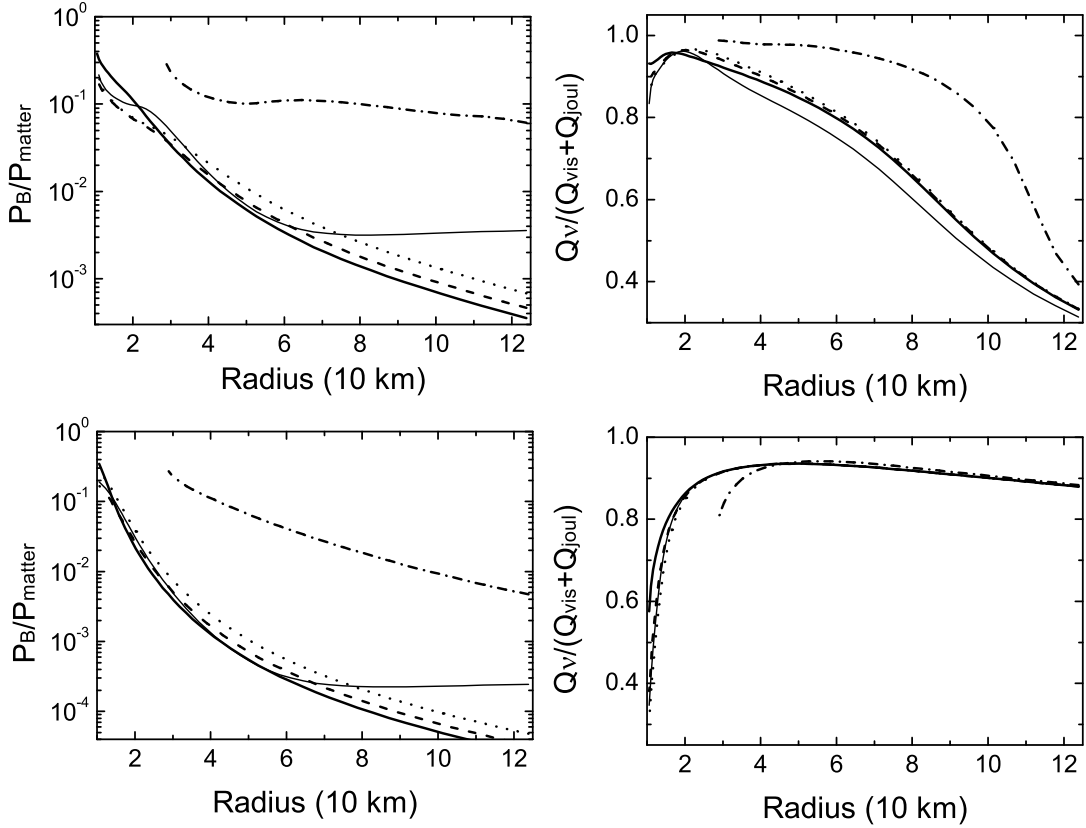}} \caption{Ratio
of magnetic pressure $P_{\rm B}$ to the accretion matter $P_{\rm
matter}$, and ratio of neutrino emission $Q_{\nu}^{-}$ to the local
heating rate $Q_{\rm vis}^{+}+Q_{\rm joul}^{+}$ as functions of disk radius.
Upper two panels for the accretion rate $M_{\odot}=0.1 M_{\odot}$
s$^{-1}$. Open disk field with $B_{0}=10^{16}$ G, $\beta=0.2, s=1$
(\textit{thick solid line}), $\beta=0.6, s=1$ (\textit{thick dashed
line}), $\beta=1, s=1$ (\textit{thick dotted line}), $B_{0}=10^{17}$
G, $\beta=1, s=1$ (\textit{thick dot-dashed line}) and closed field
with $\beta=1, s=1, P_{r}=0.005$ s (\textit{thin solid line}).
Bottom two panels for the accretion rate $M_{\odot}=1 M_{\odot}$
s$^{-1}$, and the lines are the same as the upper panels.}
\end{figure}

\newpage
\begin{figure}
\resizebox{\hsize}{!} {\includegraphics{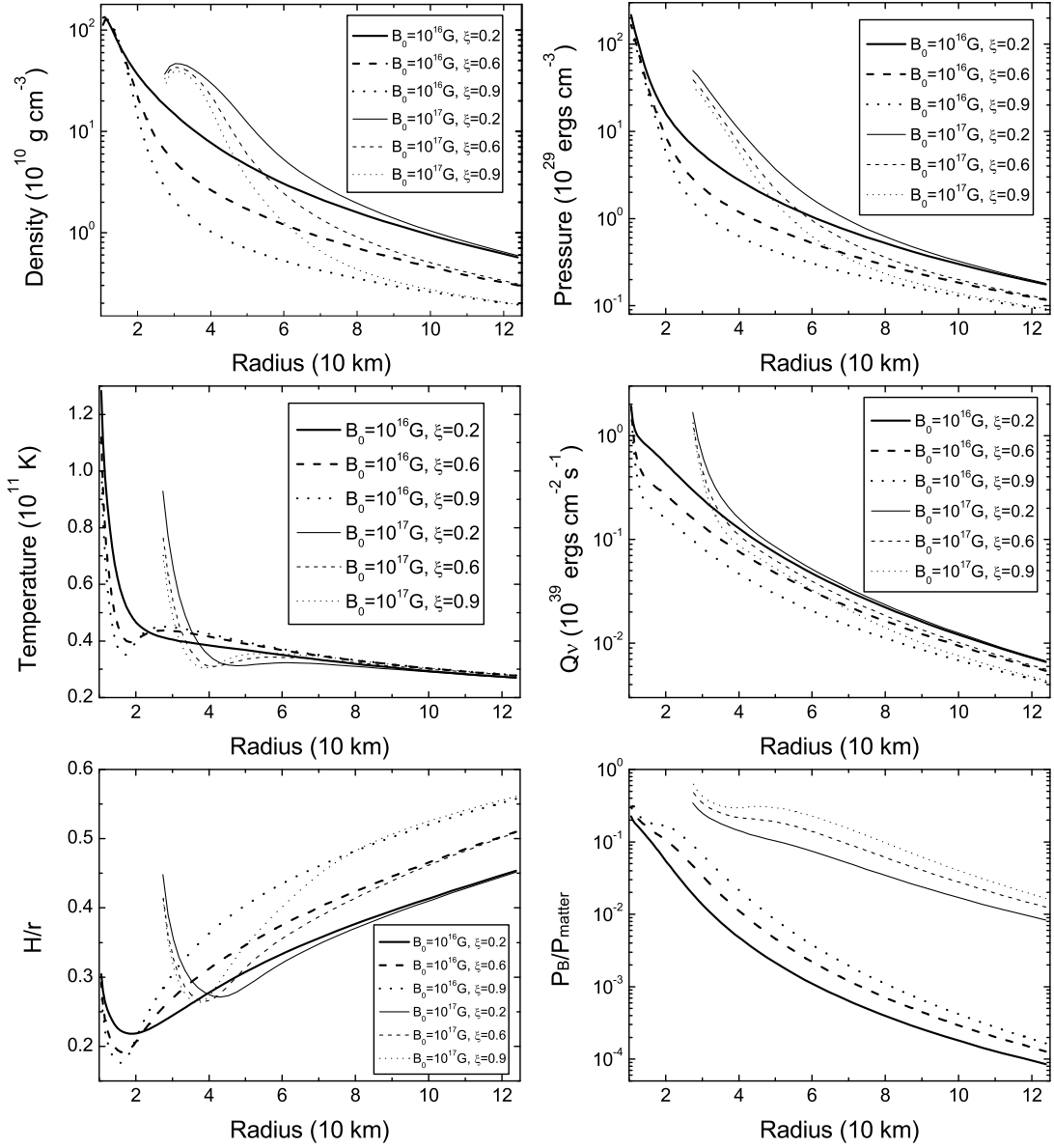}} \caption{Disk
structure and luminosity for three values of outflow index $\xi=0.2,
0.6, 0.9$ with the open disk field $\beta=0.6, s=1$, $B_{0}=10^{16},
10^{17}$ G. We give an initial accretion rate $\dot{M}=0.5M_{\odot}$
s$^{-1}$ at the outer radius $r=124$ km.}
\end{figure}
\newpage
\begin{figure}
\resizebox{\hsize}{!} {\includegraphics{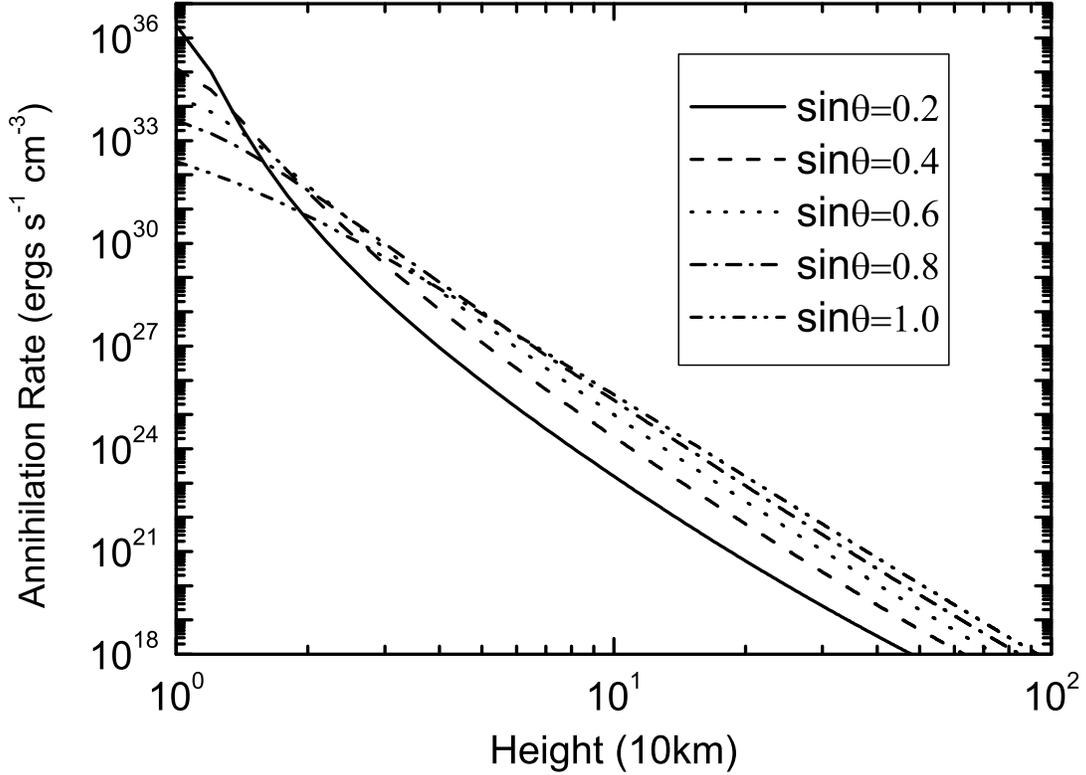}} \caption{The
neutrino annihilation rate $l_{\nu\bar{\nu}}$ (erg s$^{-1}$
cm$^{-3}$) along the stellar magnetic pole as a function of height,
where the values of $\textrm{sin}\theta\simeq\sqrt{r_{*}/r_{A}}$ are
taken as $\textrm{sin}\theta=0.2,0.4,0.6,0.8,1.0$,
$\Delta\Omega/2\pi=10^{-2}$, accretion rate $\dot{M}=0.5M_{\odot}$
s$^{-1}$, and the energy release efficiency $\epsilon=0.5$. The
integrated annihilation energy along the pole is
$\int_{r_{*}}^{\infty}
l_{\nu\bar{\nu}}\textrm{d}z=4.95\times10^{41},3.31\times10^{40},
5.92\times10^{39},1.31\times10^{38}$ and $9.21\times10^{37}$ erg
s$^{-1}$ cm$^{-2}$ for the value of
$\textrm{sin}\theta=0.2,0.4,0.6,0.8$ and $1.0$ respectively.}
\end{figure}
\newpage
\begin{figure}
\resizebox{\hsize}{!} {\includegraphics{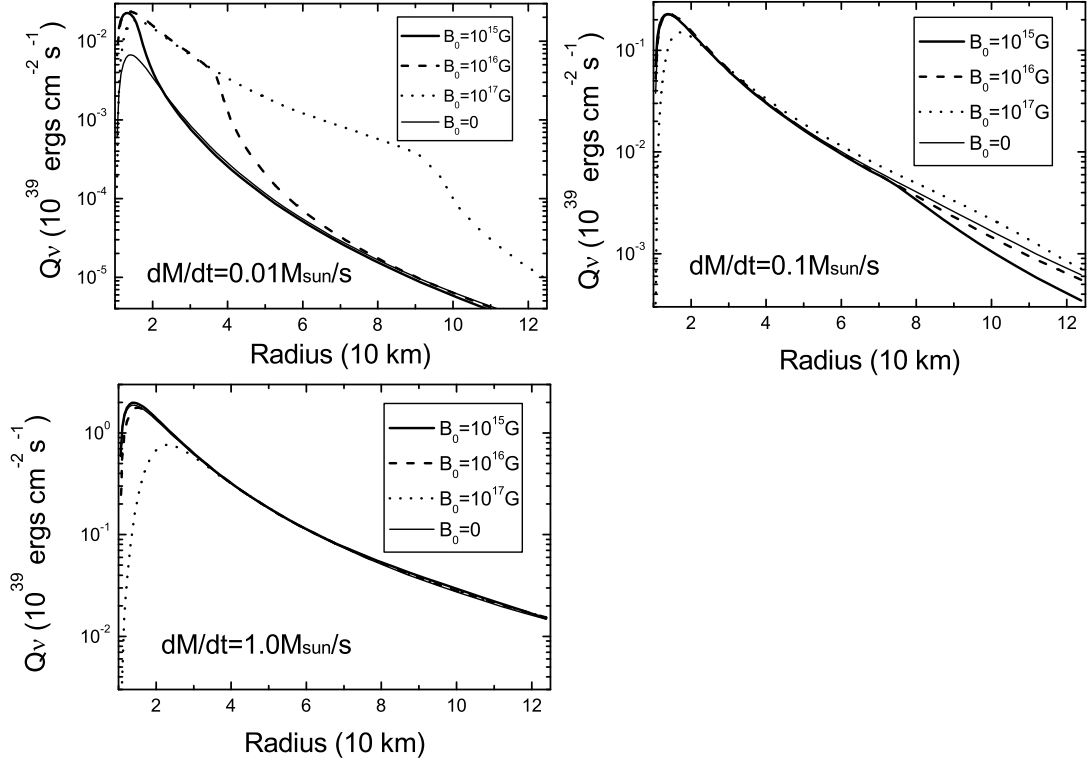}}
\caption{Neutrino luminosity for accretion rate $\dot{M}=0.01,0.1$
and 1 $M_{\odot}$ s$^{-1}$ and magnetar surface field
$B_{0}=10^{15},10^{16}$ and $10^{17}$ G without Joule dissipation in
the disk. The open field model is adopted with $\beta=0.6$ and
$s=1$.}
\end{figure}

\end{document}